\documentclass[manuscript]{aastex}
\usepackage{natbib}
\usepackage{amsmath}
\usepackage{textcomp}
\usepackage[colorlinks=true,citecolor=blue]{hyperref}
\usepackage{graphicx,color}
\usepackage{dcolumn}
\usepackage{times}
\usepackage{amssymb}
\newcommand{\Msun}{\ensuremath{M_{\odot}}}
\newcommand{\lum}{erg\,s$^{-1}$}
\newcommand{\fermi}{{\it Fermi}}
\newcommand{\phflux}{\mbox{${\rm \, ph \,\, cm^{-2} \, s^{-1}}$}}
\newcommand{\ergflux}{\mbox{${\rm \, erg \,\, cm^{-2} \, s^{-1}}$}}
\slugcomment{ApJ accepted}
\shorttitle{Bright GeV Flare of 3C 279}
\shortauthors{Paliya et al.}
\begin{document}
\title{MULTI-WAVELENGTH OBSERVATIONS OF 3C 279 DURING THE EXTREMELY BRIGHT GAMMA-RAY FLARE IN 2014 MARCH$-$APRIL}
\author{Vaidehi S. Paliya$^{1,\,2}$, S. Sahayanathan$^{3}$, and C. S. Stalin$^{1}$} 
\affil{$^1$Indian Institute of Astrophysics, Block II, Koramangala, Bangalore-560034, India}
\affil{$^2$Department of Physics, University of Calicut, Malappuram-673635, India}
\affil{$^3$Astrophysical Sciences Division, Bhabha Atomic Research Centre, Mumbai-400085, India}
\email{vaidehi@iiap.res.in}

\begin{abstract}
The well studied blazar 3C 279 underwent a giant $\gamma$-ray outburst in 2014 March$-$April. The measured $\gamma$-ray flux (1.21 $\pm$ 0.10 $\times$ 10$^{-5}$ \phflux~in 0.1$-$300 GeV energy range) is the highest detected from 3C 279 by \fermi~Large Area Telescope. Hour scale $\gamma$-ray flux variability are observed, with a flux doubling time as short as 1.19 $\pm$ 0.36 hours detected during one flare. The $\gamma$-ray spectrum is found to be curved at peak of the flare suggesting low probability of detecting very high energy (VHE; E $>$ 100 GeV) emission, which is further confirmed by the Very Energetic Radiation Imaging Telescope Array System observations. The $\gamma$-ray flux increased by more than an order in comparison to low activity state and the flare consists of multiple sub-structures having fast rise and slow decay profile. The flux enhancement is seen in all the wavebands though at a lesser extent compared to $\gamma$-rays. During the flare, a considerable amount of the kinetic jet power gets converted to $\gamma$-rays and the jet becomes radiatively efficient. A one zone leptonic emission model is used to reproduce the flare and we find increase in the bulk Lorentz factor as a major cause of the outburst. From the observed fast variability, lack of VHE detection, and the curved $\gamma$-ray spectrum, we conclude that the location of the emission region cannot be far out from the broad line region (BLR) and contributions from both BLR and torus photons are required to explain the observed $\gamma$-ray spectrum.
\end{abstract}


\section{Introduction}\label{sec:intro}
Blazars are a peculiar class of active galactic nuclei (AGN) with powerful relativistic jets aligned close to the line of sight to the observer \citep{1995PASP..107..803U}. Because of the small inclination angle, the emission from their jet is relativistically amplified. The emitted radiation, predominantly by non-thermal emission processes, is highly luminous and show rapid variations at all observed bands. Blazars are classified as flat spectrum radio quasars (FSRQs) and BL Lac objects based on the rest frame equivalent width (EW) of their broad optical emission lines, with FSRQs having EW $> 5$ \AA~\citep{1991ApJS...76..813S,1991ApJ...374..431S}. However, \citet{2009MNRAS.396L.105G} have proposed a new classification based on the broad line region (BLR) luminosity in units of Eddington luminosity, with FSRQs having higher BLR luminosity ($L_{\rm BLR}$/$L_{\rm Edd}$ $>$ 5 $\times$ 10$^{-4}$). Both classes share many common properties, such as flat radio spectra ($\alpha_r~< 0.5;~\rm{S_{\nu}}~\propto~\rm{\nu}^{-\alpha}$) at GHz frequencies, rapid flux and polarization variations \citep{1995ARA&A..33..163W,2005A&A...442...97A} and exhibit superluminal patterns at radio wavelengths \citep{2005AJ....130.1418J}.

The broadband spectral energy distribution (SED) of blazars has two broad peaks, the first between mm and soft X-ray wavelengths, and the second in the MeV$-$GeV range. In general, FSRQs exhibit lower peak energies and higher bolometric luminosities than BL Lac objects \citep{1998MNRAS.299..433F}. Various models have been proposed to explain the broadband emission from blazars. The origin of the low energy peak is understood to be associated with synchrotron radiation from relativistic electrons, whereas the high energy peak in the SED can be explained by the inverse-Compton (IC) scattering of synchrotron photons from the jet \citep[synchrotron self Compton or SSC,][]{1981ApJ...243..700K,1985ApJ...298..114M,1989ApJ...340..181G}. Alternatively, the seed photons for IC scattering can be external to the jet \citep[external Compton or EC,][]{1987ApJ...322..650B,1989ApJ...340..162M,1992A&A...256L..27D}. The plausible reservoir of seed photons for EC can be the accretion disk \citep{1993ApJ...416..458D,1997A&A...324..395B}, the BLR \citep{1994ApJ...421..153S,1996MNRAS.280...67G}, and the dusty torus \citep{2000ApJ...545..107B,2008MNRAS.387.1669G}. On contrary, the presence of the high energy peak is also attributed to hadronic processes initiated by relativistic protons co-accelerated with the electrons \citep[e.g.,][]{2003APh....18..593M,2013ApJ...768...54B}. 

The quasar 3C 279 \citep[$z$ = 0.536;][]{1965ApJ...142.1667L} is one of the first $\gamma$-ray emitting blazars discovered by the Energetic Gamma-Ray Experiment Telescope onboard the Compton Gamma-Ray Observatory \citep{1992ApJ...385L...1H}. This is also the first FSRQ detected in very high energy (VHE, E $>$ 100 GeV) $\gamma$-rays by the Major Atmospheric Gamma-ray Imaging Cherenkov (MAGIC) telescopes \citep{2008Sci...320.1752M}, thus making it one of the farthest known VHE emitters to date \citep[see also][for recent findings]{2014ATel.6349....1M}. It is strongly variable over the entire electromagnetic spectrum \citep[e.g.,][]{2012ApJ...754..114H} with the $\gamma$-ray flux varying over two orders of magnitude, from $\sim$ 10$^{-7}$ to $\sim$ 10$^{-5}$ \phflux~above 100 MeV \citep{1994ApJ...435L..91M,1998ApJ...497..178W}. An intense multi-wavelength monitoring of 3C 279 during a $\gamma$-ray flare in 2009 led to the discovery of the change in optical polarization in conjunction with the $\gamma$-ray flare \citep{2010Natur.463..919A}. This source is bright in the hard X-ray band and included in the 70 month Swift-Burst Alert Telescope (BAT) catalog \citep{2013ApJS..207...19B}. A multi-frequency study covering the first two years of {\it Fermi Gamma-ray Space Telescope} mission has found the $\gamma$-rays to lead the optical emission by $\sim$10 days \citep{2012ApJ...754..114H}. The detection of an orphan X-ray flare with no clear counterpart in other wavebands is also reported by them. Interestingly, there are observations of inconsistent patterns of correlation over various energy bands shown by 3C 279 \citep[e.g.,][]{2008ApJ...689...79C}. At the radio wavelengths, 3C 279 exhibits a compact core and Very Long Baseline Array observations revealed superluminal patterns with apparent speed of (20.6 $\pm$ 0.8)c \citep{2013AJ....146..120L}. Moreover, radio studies have also estimated the bulk Lorentz factor and viewing angle of the jet flow as $\Gamma_{\rm j}$ = 15.5 $\pm$ 2.5 and $\Theta_{\rm j}$ = 2$^{\circ}$.1 $\pm$ 1$^{\circ}$.1 \citep{2004AJ....127.3115J,2005AJ....130.1418J}.

Recently, 3C 279 was detected in an exceptionally high activity state \citep{2014ATel.6036....1C} by the Large Area Telescope (LAT) onboard \fermi~\citep[hereafter \fermi-LAT;][]{2009ApJ...697.1071A}. We denote the period 2014 March 25 to 2014 April 13 (MJD 56,741$-$56,760) as high activity period. During this period, some of the brightest $\gamma$-ray flares were observed from 3C 279 since the beginning of \fermi-LAT operation. A special 350 ksec target of opportunity (ToO) observation was approved (between MJD 56,747$-$56,755) during which \fermi-LAT monitored this source in pointing mode, other than its normal survey mode operation. This flaring event was simultaneously monitored at low frequencies by the {\it Swift} satellite \citep{2004ApJ...611.1005G} and many ground based observational facilities. In this work, motivated by the availability of near simultaneous multi-wavelength data, we study this exceptional $\gamma$-ray outburst in detail. We also discuss the implications of our findings to constrain the location of the $\gamma$-ray emission region during the flare using a multi-wavelength approach. In Section~\ref{sec:data_red}, we report the details of the data reduction procedure and the results are presented in Section~\ref{sec:results}. We discuss our findings in Section~\ref{sec:dscsn} and conclude in Section~\ref{sec:summary}. Throughout the work, we adopt a $\Lambda$CDM cosmology with the Hubble constant $H_0=71$~km~s$^{-1}$~Mpc$^{-1}$, $\Omega_m = 0.27$, and $\Omega_\Lambda = 0.73$.

\section{Multiwavelength observations and Data Reduction}\label{sec:data_red}
\subsection{{\it Fermi}-Large Area Telescope Observations}\label{subsec:fermi}

The Fermi-LAT data used in this work were collected covering the period of the outburst (MJD 56,741$-$56,760) and a separate low activity period (MJD 55,300$-$55,400). The standard data analysis procedure as mentioned in the \fermi−LAT documentation\footnote{http://fermi.gsfc.nasa.gov/ssc/data/analysis/documentation/} is adopted. Events belonging to the energy range 0.1$-$300 GeV and SOURCE class are used. To select good time intervals, a filter ``\texttt{DATA$\_$QUAL$>$0}'', \&\& ``\texttt{LAT$\_$CONFIG==1}'' is used and a cut of 100$^{\circ}$ is also applied on the zenith angle to avoid contamination from the Earth limb $\gamma$-rays. We consider the recently released galactic diffuse emission component gll\_iem\_v05\_rev1.fits and an isotropic component iso\_source\_v05\_rev1.txt as background models\footnote{http://fermi.gsfc.nasa.gov/ssc/data/access/lat/BackgroundModels.html}, whose normalization parameters are left free to vary during the fitting. The unbinned likelihood method included in the pylikelihood library of {\tt Science Tools (v9r33p0)} and the post-launch instrument response functions P7REP\_SOURCE\_V15 are used for the analysis. Significance of the $\gamma$-ray signal is computed by means of the maximum likelihood (ML) test statistic TS = 2$\Delta \log (\mathcal{L}$) where $\mathcal{L}$ represents the likelihood function between models with and without a point source at the position of the source of interest. All the sources lying within 10$^{\circ}$ region of interest (ROI) centered at the position of 3C 279 and defined in the second \fermi-LAT catalog \citep{2012ApJS..199...31N}, are included in the analysis. All the parameters except the scaling factor of the sources within the ROI are allowed to vary during the likelihood fitting. Additionally, we also include the sources lying within 10$^{\circ}$ to 15$^{\circ}$ from the center of the ROI and keep their parameters fixed to the 2FGL catalog value. The $\gamma$-ray bright blazar 3C 273 lies at $\sim$10$^{\circ}$.3 from 3C 279, and we keep its spectral parameters free during the fitting. We also search for the presence of unmodeled sources within the ROI by generating the residual TS maps for the periods covered in this work. We do not find any significant unmodeled source (i.e. source with TS $>$ 25). We perform a first run of the ML analysis over the period of interest and remove all the sources with TS $<$ 25. This updated model is then used for further temporal and spectral analysis. Though 3C 279 is modeled by a logParabola model in the 2FGL catalog, to generate light curves, we use a power law (PL) model as the PL indices obtained from this model show smaller statistical uncertainties when compared to those obtained from complex model fits. Moreover, since we want to probe the shortest timescales (hence lower photon statistics), adopting a simple PL model is appropriate. For the time series and spectral analysis, we consider the source to be detected if TS $>$ 9 which corresponds to $\sim$3$\sigma$ detection \citep{1996ApJ...461..396M}. Bins with TS $<$ 9 and/or $\Delta F_{\gamma}/F_{\gamma} > 0.5$, where $\Delta F_{\gamma}$ is the error estimate in the flux $F_{\gamma}$, are rejected from the analysis. Primarily governed by uncertainty in the effective area, the measured fluxes have energy dependent systematic uncertainties of around 10\% below 100 MeV, decreasing linearly in log(E) to 5\% in the range between 316 MeV and 10 GeV and increasing linearly in log(E) up to 15\% at 1 TeV\footnote{http://fermi.gsfc.nasa.gov/ssc/data/analysis/LAT\_caveats.html}. All errors associated with the LAT data analysis are the 1$\sigma$ statistical uncertainties, unless specified.

\subsection{{\it Swift} Observations}\label{subsec:swift}
Throughout the flaring period, the {\it Swift} satellite has monitored 3C 279 almost regularly using all the three instruments onboard it. However, due to the poor sensitivity of BAT \citep{2005SSRv..120..143B}, it is not possible to extract signal over the short time periods covered in this work. On the other hand, it is significantly detected by the X-Ray Telescope \citep[XRT;][]{2005SSRv..120..165B} as well as by the UltraViolet Optical Telescope \citep[UVOT;][]{2005SSRv..120...95R}.

The XRT data are first processed with the XRTDAS software package (v.3.0.0) available within HEASOFT package (6.16). Event files are cleaned and calibrated using standard procedures ({\tt xrtpipeline v.0.13.0}) with the calibration database updated on 2014 November 7. Standard grade selections of 0$-$12 in the photon counting mode are used. Energy spectrum is extracted from the summed event files. Since at the peak of the $\gamma$-ray flare, 3C 279 was extremely bright in the 0.3$-$10 keV band, we select annular regions centered at the source position to extract the source and the background spectra. This is required to avoid possible pile up effects. Inner and outer radii of the source region are chosen as 5$^{\prime\prime}$ and 65$^{\prime\prime}$ respectively, while the background spectra are extracted from an annular region of inner and outer radii of 130$^{\prime\prime}$ and 230$^{\prime\prime}$ respectively. Selection of this particular choice of radii of annular regions is based on the task {\tt xrtgrblc v.1.6 4}\footnote{http://heasarc.gsfc.nasa.gov/lheasoft/ftools/headas/xrtgrblc.html} \citep[see also][]{2013ApJS..207...28S}. Exposure maps are combined using XIMAGE and ancillary response files are generated using the task {\tt xrtmkarf}. Source spectra are binned to have at least 20 counts per bin, using the task {\tt grppha}. Spectral fitting is done using Xspec \citep{1996ASPC..101...17A}. An absorbed power law \citep[N$_{\rm H}$ = 2.05 $\times$ 10$^{20}$ cm$^{-2}$;][]{2005A&A...440..775K} is used for fitting and the uncertainties are calculated at 90\% confidence level.

{\it Swift}-UVOT observations are integrated using {\tt uvotimsum} and the parameters are extracted using the task {\tt uvotsource}. Source region is selected as a circle centered at the source position and of 5$^{\prime\prime}$ radius, while background is chosen from a nearby source-free circular region of 1$^{\prime}$ radius. Observed magnitudes are corrected for galactic extinction following \citet{2011ApJ...737..103S} and converted to flux units using the zero point magnitudes and conversion factors of \citet{2011AIPC.1358..373B}.

\subsection{SMARTS Observations}\label{subsec:smarts}
A sample of $\gamma$-ray emitting AGN discovered by \fermi-LAT is being monitored by Small and Moderate Aperture Research Telescope System (SMARTS) at the Cerro Tololo Inter-American Observatory located at Chile. Optical and near-infrared (IR) data from SMARTS are routinely available in B, V, R, J, and K bands. More details on data acquisition, reduction and calibration can be found in \citet{2012ApJ...756...13B}. For this work, we collected the publicly available data of 3C 279. The data in all the filters are corrected for galactic extinction following \citet{2011ApJ...737..103S} and then converted to flux units using the zero points of \citet{1998A&A...333..231B}.

\subsection{Steward Observatory Monitoring}\label{subsec:steward}
Optical photometric, spectrophotometric, and spectropolarimetric observations of \fermi-LAT detected blazars are being carried out at the Steward observatory at the university of Arizona as part of its monitoring program. Details of the data reduction and calibration procedures are presented in \citet{2009arXiv0912.3621S}. We downloaded the publicly available optical photometric and polarimetric data of 3C 279. The photometric V band observations are corrected for galactic reddening \citep{{2011ApJ...737..103S}} and converted to flux units following \citet{1998A&A...333..231B}.

\section{Results}\label{sec:results}
\begin{figure*}
\hbox{
      \includegraphics[width=\columnwidth]{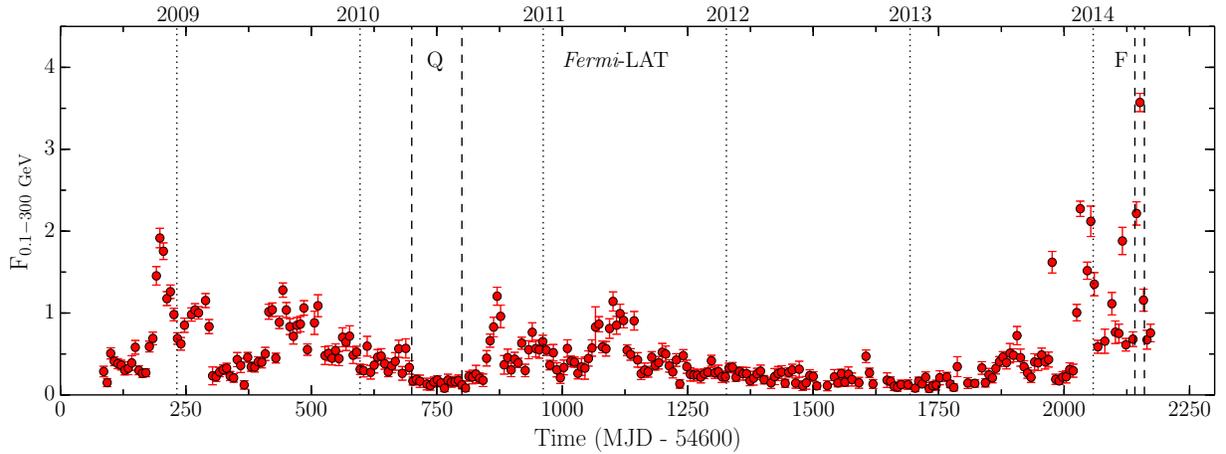}
     }
\caption{Weekly binned $\gamma$-ray light curve of 3C 279 with flux units of 10$^{-6}$ \phflux. A low activity (Q) and flaring (F) periods are shown by dashed lines, whereas dotted lines represent the beginning of the new years.}\label{fig:week_lc}
\end{figure*}

We select the period 2014 March 25 to 2014 April 13 (MJD 56,741$-$56,760) for a detailed study of this brightest $\gamma$-ray flare observed from 3C 279 by \fermi-LAT. For comparison, we also select a $\gamma$-ray low activity state, 2010 April 14 to 2010 July 23 (MJD 55,300$-$55,400). These selected periods are shown with the symbols $F$ and $Q$, representing flaring and low activity state respectively in Figure~\ref{fig:week_lc}, where we present the weekly binned $\gamma$-ray light curve of 3C 279 since the beginning of \fermi-LAT operation.

\subsection{Multi-band Temporal Variability}\label{subsec:mw_var}
In Figure~\ref{fig:mw_lc}, we show the multi-band light curves of 3C 279 from $\gamma$-rays to IR as well as optical polarization measurements, covering the period of high activity. In this plot, \fermi-LAT data points are one day binned, whereas, the observations in other wavebands are one point per observation id. We divide the period of high activity into three sub-periods; Flare 1 (MJD 56,741$-$56,749), Flare 2 (MJD 56,749$-$56,755), and post$-$flare (MJD 56,755$-$56,760). These sub-periods are also selected taking into account the availability of near simultaneous observations in other energy bands. From the low cadence multi-band light curves in Figure~\ref{fig:mw_lc}, the flux variations at different wavelengths appear to be correlated. However, the presence or absence of time delays between different bands could not be statistically ascertained due to the less number of data points.

The multi-wavelength variability amplitudes of 3C 279 during the period of high activity are analyzed using the fractional rms variability amplitude parameter $F_{\rm var}$ \citep[e.g.,][]{2003MNRAS.345.1271V}. It is calculated as follows
\begin{equation} 
F_{\rm var}={(S^2-\Delta^2)^{1/2}\over\langle r\rangle}
 \end{equation} 
\noindent
where $S^2$ is the sample variance, $\langle r\rangle$ the unweighted mean count rate, and $\Delta^2$ the mean square value of the uncertainties associated with each individual count rate. The error on $F_{\rm var}$ is computed following \citet{2003MNRAS.345.1271V}
\begin{equation}
\sigma_{F_{\rm var}}= \sqrt{\left( \sqrt{\frac{1}{2N}}\cdot\frac{\Delta^2}{\langle r\rangle^2 F_{\rm var}} \right)^2+\left( \sqrt{\frac{\Delta^2}{N}}\cdot\frac{1}{\langle r\rangle}   \right)^2}
\end{equation}
\noindent
where $N$ is the number of data points. The maximum F var is found for the $\gamma$-ray band and it decreases with frequency (see Table~\ref{tab:f_var}), a trend generally found in blazars \citep[e.g.,][]{2005ApJ...629..686Z,2010ApJ...712..405V}. The unusual high $F_{\rm var}$ for the K-band light curve could be due to the presence of two adjacent points where flux is varying abruptly (MJD 56,750 and 56,751). This is possibly due to bad weather conditions present during the observing run (SMARTS team, private communication).

The good $\gamma$-ray photon statistics during this exceptional flaring event permit us to search for short timescale variability by using finer time bins. We generate twelve hours, six hours, and three hours binned $\gamma$-ray light curves covering the period of high activity and show them in Figure~\ref{fig:rapid}. In this figure, black data points correspond to the observations taken during the ToO monitoring. We scan this light curve to search for short time variability using the following equation
\begin{equation}\label{eq:flux_double}
F(t) = F(t_0).2^{(t-t_0)/\tau}
\end{equation}
where $F(t)$ and $F(t_0)$ are the fluxes at time $t$ and $t_0$ respectively, and $\tau$ is the characteristic doubling/halving time scale. We also set the condition that the difference in flux at the epochs $t$ and $t_0$ is at least significant at the 3$\sigma$ level \citep{2011A&A...530A..77F}. The results of this analysis are presented in Table~\ref{tab:shortest_gamma_var}. The shortest flux doubling time is found to be 1.19 $\pm$ 0.36 hours for the flare that occurred on MJD 56,746. Moreover, along with the use of three hours binning, we also analyze the data using the time bins defined as Good Time Intervals \citep[GTI;][]{2011A&A...530A..77F}. A GTI corresponds to the shortest time interval when the LAT data can be considered `valid'\footnote{http://fermi.gsfc.nasa.gov/ssc/data/analysis/scitools/help/gtmktime.txt}. The shortest flux doubling time using this method is obtained as 1.27 $\pm$ 0.36 hours on MJD 56,746, with $\sim$4$\sigma$ significance. This is the first report of hour scale $\gamma$-ray variability detected from 3C 279 since the launch of \fermi. The highest one day averaged photon flux is found to be (6.54 $\pm$ 0.30) $\times$ 10$^{-6}$ \phflux~in the bin centered at MJD 56,750.5 and the corresponding photon index is 2.22 $\pm$ 0.04. Moreover, from the three hours binned $\gamma$-ray light curve, the peak flux and the associated photon index are found to be (1.21 $\pm$ 0.10) $\times$ 10$^{-5}$ \phflux~and 2.19 $\pm$ 0.09 respectively, again on the same day, i.e. MJD 56,750 (see Figure~\ref{fig:rapid}). This is the highest $\gamma$-ray flux measurement from 3C 279 since the beginning of \fermi-LAT mission.

The shortest 0.3$-$10 keV X-ray flux doubling time in the observed frame, estimated using equation~\ref{eq:flux_double}, is 8.11 $\pm$ 0.98 hours measured on MJD 56,752 with $\sim$9$\sigma$ confidence. This coincides with a $\gamma$-ray flare (see Figure~\ref{fig:rapid}). The highest X-ray flux is found as 4.54$^{+0.62}_{-0.49}$ $\times$ 10$^{-11}$ \ergflux~measured on MJD 56,752. The obtained photon index is hard and having a value of 1.31$^{+0.12}_{-0.13}$. This corresponds to an isotropic X-ray luminosity of 3.82 $\times$ 10$^{46}$ \lum.
\subsection{Highest Energy Gamma-ray Photon}\label{subsec:highest_gamma_photon}
To determine the energy of the highest energy photon detected from the source, we analyze the LAT data using event class CLEAN. The tool {\tt gtsrcprob} is used for this purpose. We find the highest energy photon of 13.54 GeV detected on 2014 April 3 (MJD 56,750.46209) at 2$^{\prime}$.52 away from the position mentioned in the 2FGL catalog. The probability that the highest energy photon can be associated to the location, offset by 2$^{\prime}$.52 from the position of 3C 279 mentioned in the 2FGL catalog, is 99.94\%.

\subsection{Gamma-ray Spectral Analysis}\label{subsec:gamma_spec}
We generate the $\gamma$-ray spectra for all the four periods under consideration, namely Flare 1, Flare 2, post$-$flare, and a low activity state. Analysis of the $\gamma$-ray spectral shape is done using two spectral models: power law ($dN/dE \propto E^{\Gamma_{\gamma}}$), where $\Gamma_{\gamma}$ is the photon index and logParabola ({ $dN/dE \propto (E/E_{\rm o})^{-\alpha-\beta log({\it E/E_{\rm o}})}$}, where $E_{\rm o}$ is an arbitrary reference energy fixed at 300 MeV, $\alpha$ is the photon index at $E_{\rm o}$ and $\beta$ is the curvature index which defines the curvature around the peak). To test for the presence of curvature, the test statistic of the curvature $TS_{\rm curve}$ = 2(log $\mathcal{L}$(LogParabola) $-$ log $\mathcal{L}$(power-law)), is calculated. A $TS_{\rm curve}$ having value greater than 16 suggests for the presence of significant curvature in the $\gamma$-ray spectrum \citep{2012ApJS..199...31N}. The resultant SEDs in the $\gamma$-ray band are shown in Figure~\ref{fig:gamma_spec} and the fitting parameters are given in Table~\ref{tab:gamma_spec}. Significant curvature is noticed only during the Flare 2 with $TS_{\rm curve}$ $\approx$ 30. Though at low significance, there is also a hint for the presence of curvature in the Flare 1 state ($TS_{\rm curve}$ $\approx$ 8).

\subsection{Spectral Energy Distributions}\label{subsec:sed}
\subsubsection{Model Setup}\label{subsubsec:model}
We develop a simple one zone leptonic emission model by following the procedures outlined in \citet[hereafter GT09]{2009MNRAS.397..985G} and \citet{2009ApJ...692...32D} \citep[see also][]{2008ApJ...686..181F} to interpret the broadband emission from 3C 279. The emission region is assumed to be spherical, located at a distance of $R_{\rm diss}$ from the central black hole, and filled with electrons having smooth broken power law energy distribution
\begin{equation}\label{eq:EED}
 N'(\gamma')  \, = \, N'_0\, { (\gamma'_{\rm b})^{-p} \over
(\gamma'/\gamma'_{\rm b})^{p} + (\gamma'/\gamma'_{\rm b})^{q}},
\end{equation}
 where $p$ and $q$ are the particle indices before and after the break energy ($\gamma'_{\rm b}$) respectively (primed quantities are measured in the comoving frame). The emission region size is adopted by considering it to cover the entire jet cross-section with jet semi opening angle assumed as 0.1 degree. The accretion disk is assumed to be of standard \citet{1973A&A....24..337S} type and producing a multi-temperature blackbody spectrum \citep{2002apa..book.....F}. It extends from $R_{\rm in,d}$ = 3$R_{\rm Sch}$ to $R_{\rm out,d}$ = 500$R_{\rm Sch}$ , where $R_{\rm Sch}$ is the Schwarzschild radius. Above and below the accretion disk, the presence of X-ray corona is also considered which reprocesses a fraction $f_{\rm cor}$ of the accretion disk luminosity. The inner and outer radii of the corona are assumed to be 3$R_{\rm Sch}$ and 30$R_{\rm Sch}$ respectively. The spectrum of the X-ray corona is considered to be a cut-off power law: $L_{\rm cor}(\epsilon)\propto \epsilon^{-\alpha_{\rm cor}}\exp(-\epsilon/\epsilon_{\rm c})$ (GT09). The BLR is assumed to be a spherical shell located at a distance of $R_{\rm BLR} = 10^{17} L^{1/2}_{\rm d,45}$ cm, where $L_{\rm d,45}$ is the accretion disk luminosity in units of 10$^{45}$ \lum. It reprocesses a fraction $f_{\rm BLR}$ of the accretion disk luminosity. The SED of the BLR is approximated as an isotropic blackbody peaking at the rest-frame frequency of the Lyman-$\alpha$ line \citep{2008MNRAS.386..945T}. The dusty torus, for simplicity, is assumed to be a thin spherical shell located at a distance $R_{\rm torus} = 10^{18} L^{1/2}_{\rm d,45}$ cm, reprocessing a fraction $f_{\rm torus}$ of the accretion disk radiation in the infrared. The spectrum of the torus is assumed as a blackbody with temperature $T_{\rm torus}$ = $\epsilon_{\rm peak}m_{\rm e}c^2/3.93k$, where $\epsilon_{\rm peak}$ is the dimensionless peak photon energy and $k$ is the Boltzmann constant. Following GT09, we calculate the relative contribution of these emissions with respect to the distance from the central black hole. Moreover, the synchrotron and SSC spectra are computed using the prescriptions of \citet{2008ApJ...686..181F}. The external Compton emissions are calculated following GT09 \citep[see also][]{2009ApJ...692...32D,2009herb.book.....D}. Finally, kinetic power of the jet is calculated by assuming protons to be cold, contributing only to the inertia of the jet, and having equal number density to that of the relativistic electrons \citep[e.g.,][]{2008MNRAS.385..283C}. To model the SEDs, we start with a plausible set of parameters which are then constrained by reproducing the observed fluxes at different energies.

\subsubsection{SED Modeling Results}\label{subsubsec:sed_results}
We generate the broadband SEDs of 3C 279 during a low activity period and three sub-periods covering the flaring event. The fluxes over each of the four time intervals are averaged and the derived values are given in Table~\ref{tab:sed_flux}, except for \fermi-LAT data which are presented in Table~\ref{tab:gamma_spec}. The broadband SEDs are reproduced using the model presented in Section~\ref{subsubsec:model} and with the following assumptions: the spectral shape of the X-ray corona is assumed to be flat ($\alpha_{\rm cor}$ = 1) and the high energy cut-off is fixed at 150 keV (GT09). Fractions of the accretion disk luminosity reprocessed by the X-ray corona, the BLR, and the dusty torus are adopted as 0.3, 0.1, and 0.5 respectively.

As discussed by \citet{2010MNRAS.402..497G}, a powerful diagnostic to constrain the accretion disk luminosity and the black hole mass is through modeling of the accretion disk spectrum over the optical$-$UV part of the SED, provided the optical$-$UV bump (the big blue bump) is visible. Since the optical$-$UV spectrum of 3C 279 is dominated by non-thermal spectrum \citep[e.g.,][]{2012ApJ...754..114H}, we start with the modeling of the low activity state where the probability of detecting thermal emission from the accretion disk is high. The optical-UV spectrum during the low activity state shows a turnover at high energies (Figure~\ref{fig:sed_fit}), though it is not prominent.  We attribute this excess to the accretion disk radiation. The observed optical-UV spectrum is then reproduced by a combination of synchrotron and the accretion disk emissions. Accordingly, the derived accretion disk luminosity and the black hole mass are 1 $\times$ 10$^{45}$ \lum~and 3 $\times$ 10$^8$ \Msun~respectively. These values are quite in agreement with that obtained in earlier studies \citep[2 $\times$ 10$^{45}$ \lum, 3 $\times$ 10$^{8}$ \Msun;][]{1999ApJ...521..112P,2002ApJ...579..530W}. Using the obtained accretion disk luminosity and the black hole mass, we model the SEDs covering the flaring period. The model spectra due to different emission mechanisms along with the observed fluxes are shown in Figure~\ref{fig:sed_fit} and the relevant parameters are given in Table~\ref{tab:sed}.

The size of the emission region is obtained as $R_{\rm blob}$ $\sim$1 $\times$ 10$^{16}$ cm, constrained from the SED modeling. This corresponds to a variability time ($t_{\rm v} = R_{\rm blob}$ (1 + $z$)/$\delta c$; where Doppler factor $\delta$ = 19 taken from the SED modeling) of $\sim$8 hours in the observer’s frame. The similarity of the variability time deduced from the SED modeling with that obtained using the shortest X-ray flux doubling timescale is striking. However, we find the shortest $\gamma$-ray flaring time as small as $\sim$1 hour (Table~\ref{tab:shortest_gamma_var}). This hints for the existence of multiple $\gamma$-ray emitting sub-structures within a larger emission region responsible for the flux enhancement in all the wavebands. These smaller regions could be responsible for the observed fast $\gamma$-ray variations. This is also supported from the observed three hours binned $\gamma$-ray light curve where many short time flaring events are seen (Figure~\ref{fig:rapid}). Since we generate the SEDs of the source by averaging the fluxes over larger time intervals, the results of the SED modeling obtained are more likely the representation of the average characteristics of 3C 279 during the various activity states considered here.

\section{Discussion}\label{sec:dscsn}
The giant $\gamma$-ray outburst of 3C 279 in 2014 March$-$April together with the availability of near-simultaneous coverage at other wavelengths, has made it possible to study this peculiar event in detail.

\begin{figure*}
\hbox{
      \includegraphics[width=\columnwidth]{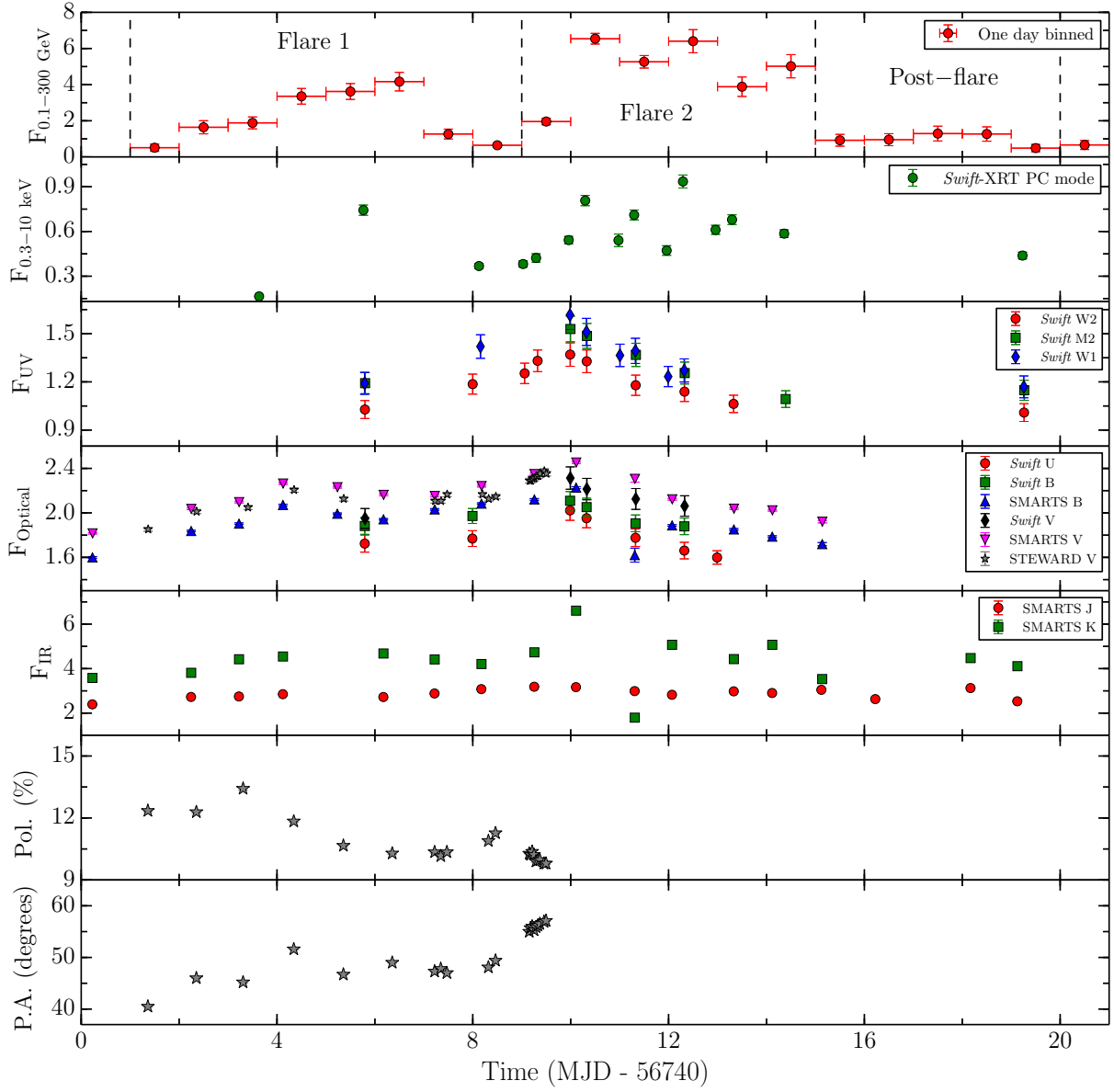}
     }
\caption{Multi-frequency light curve of 3C 279 covering the period of high activity. \fermi-LAT and {\it Swift}-XRT data points are in units of 10$^{-6}$ \phflux~and counts s$^{-1}$ respectively. UV, optical and IR fluxes have units of 10$^{-11}$ \ergflux. See text for details.}\label{fig:mw_lc}
\end{figure*}

\begin{figure*}
\hbox{
      \includegraphics[width=\columnwidth]{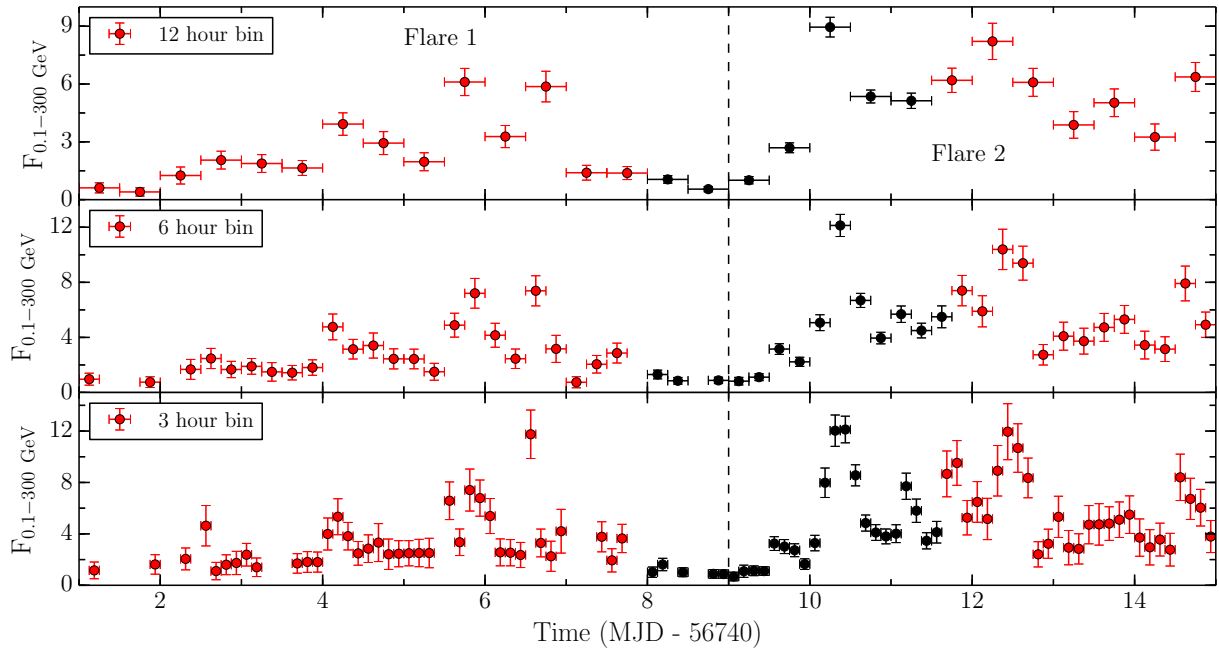}
     }
\caption{\fermi-LAT light curve of 3C 279 around the period of high activity, binned in the interval of 12 hours, 6hours, and 3 hours (upper, middle, and lower panel respectively). Black data points represents the \fermi~ToO observations. The fluxes are in units of 10$^{-6}$ \phflux.}\label{fig:rapid}
\end{figure*}

\begin{figure*}
\hbox{
      \includegraphics[width=\columnwidth]{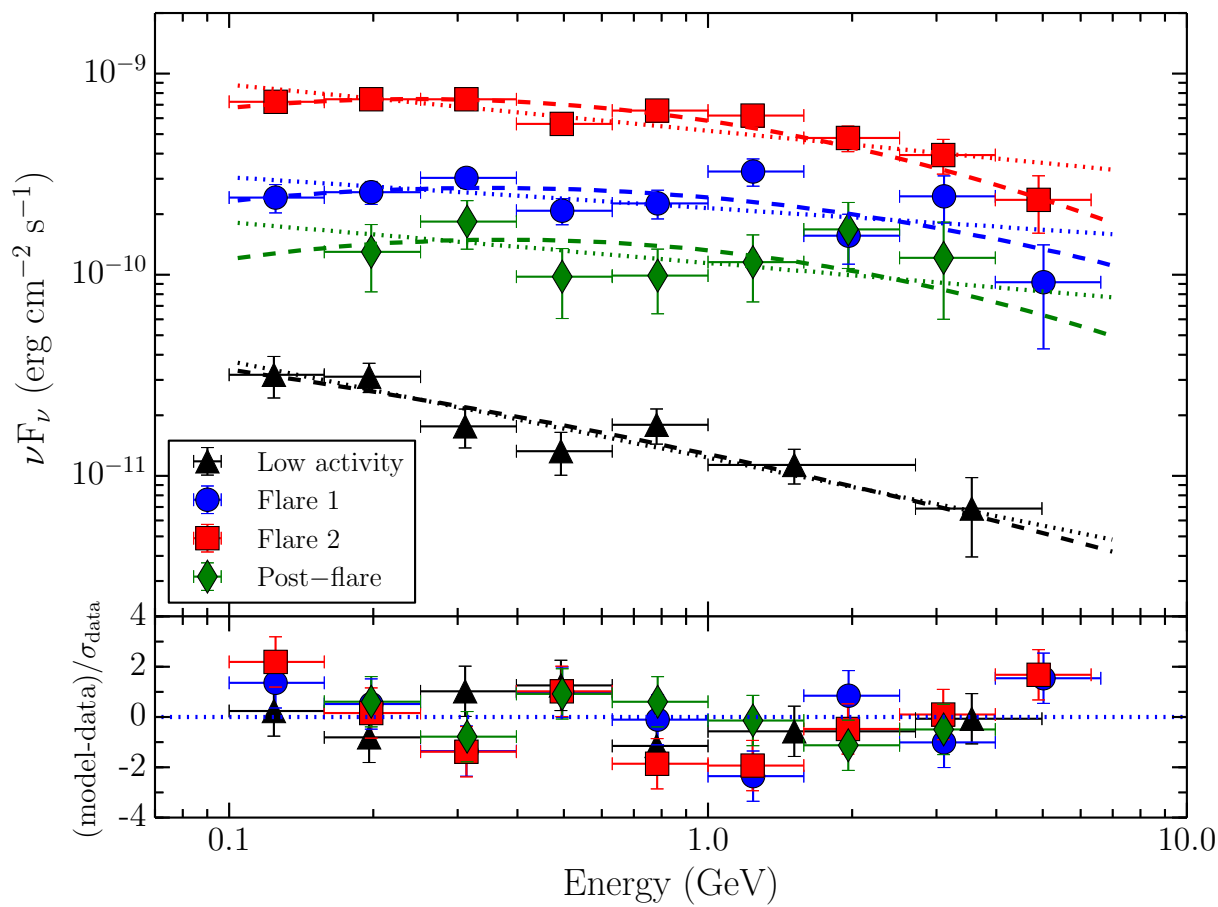}
     }
\caption{\fermi-LAT SEDs of 3C 279 during different activity states as defined in Table~\ref{tab:gamma_spec}. Power law and logParabola models are shown with dotted and dashed lines respectively. Horizontal error bars correspond to energy ranges of each bin, whereas vertical bars represent 1$\sigma$ statistical errors. The residuals in the lower panel refers to the power law model.}\label{fig:gamma_spec}
\end{figure*}

\begin{figure*}
\begin{center}
\hbox{
      \includegraphics[width=8cm]{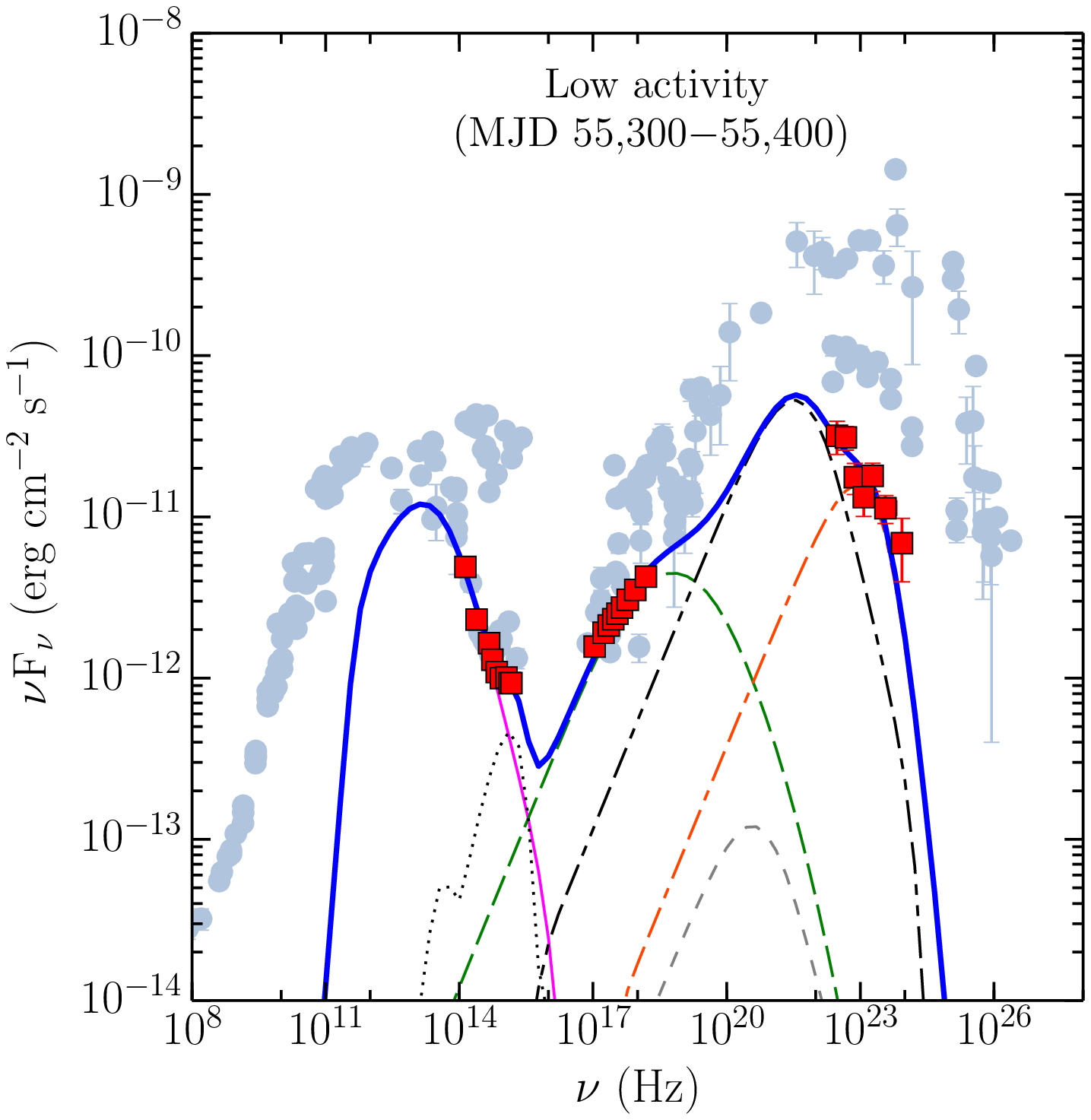}
      \includegraphics[width=8cm]{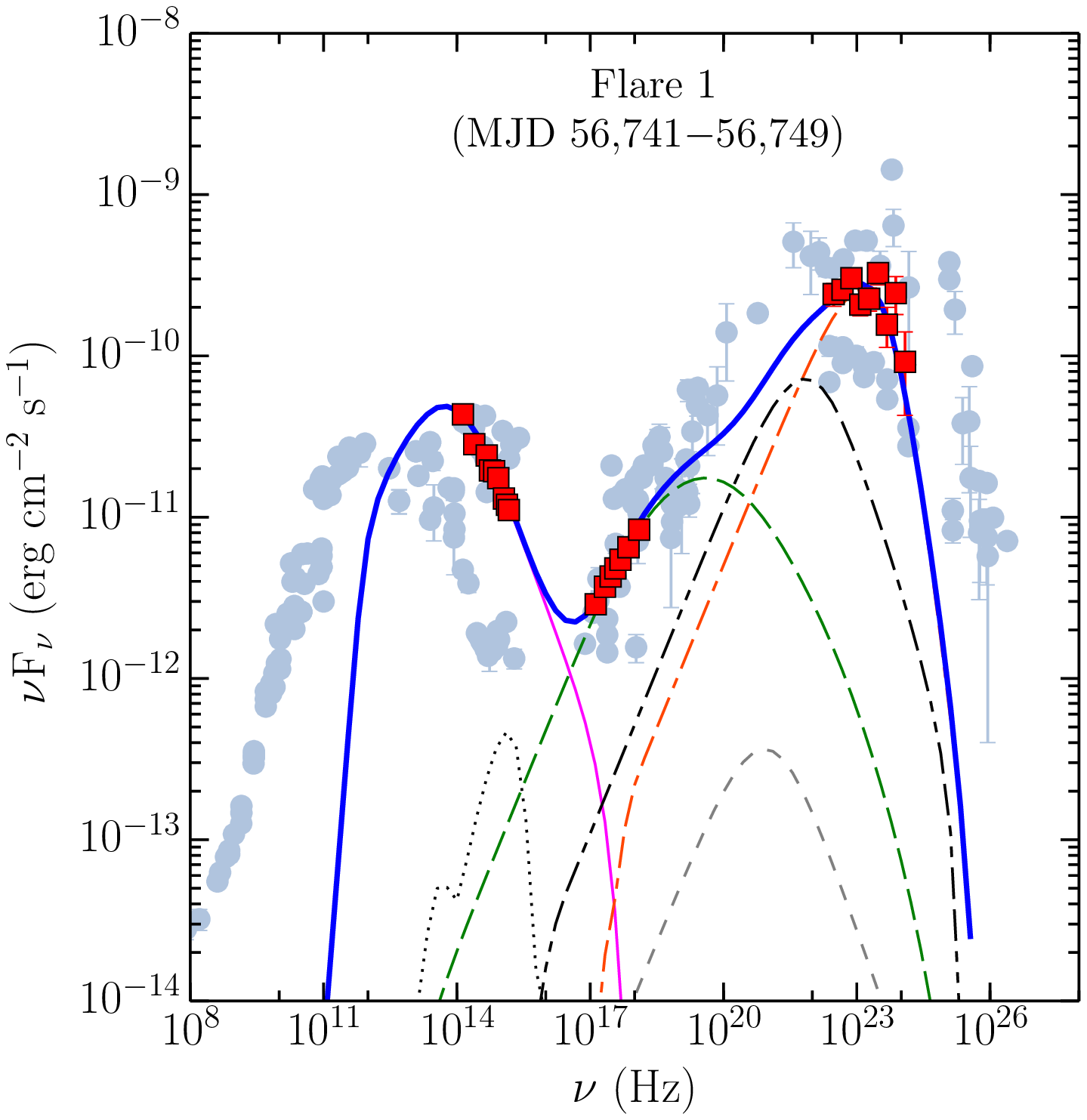}
     }
\hbox{
      \includegraphics[width=8cm]{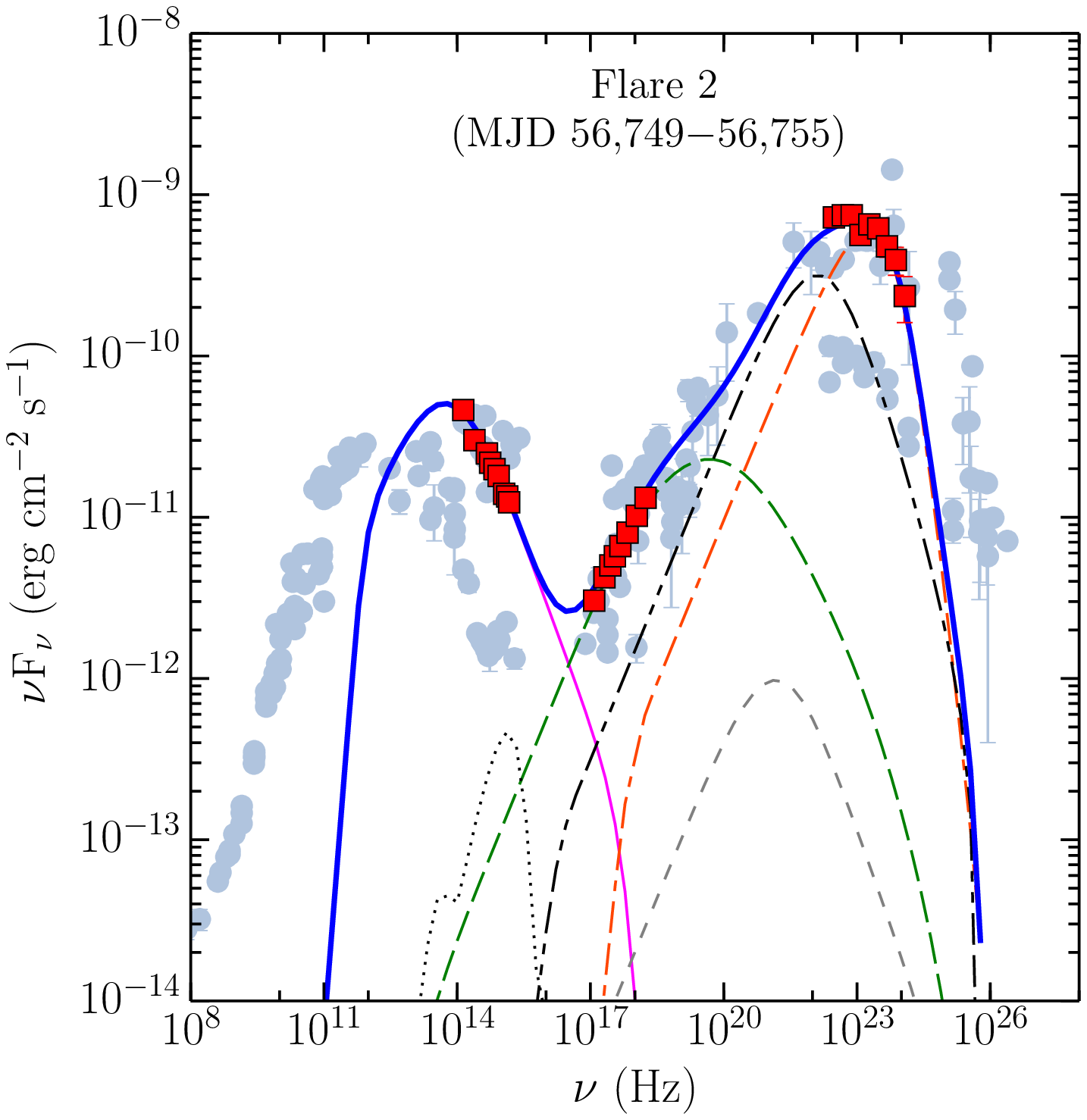}
      \includegraphics[width=8cm]{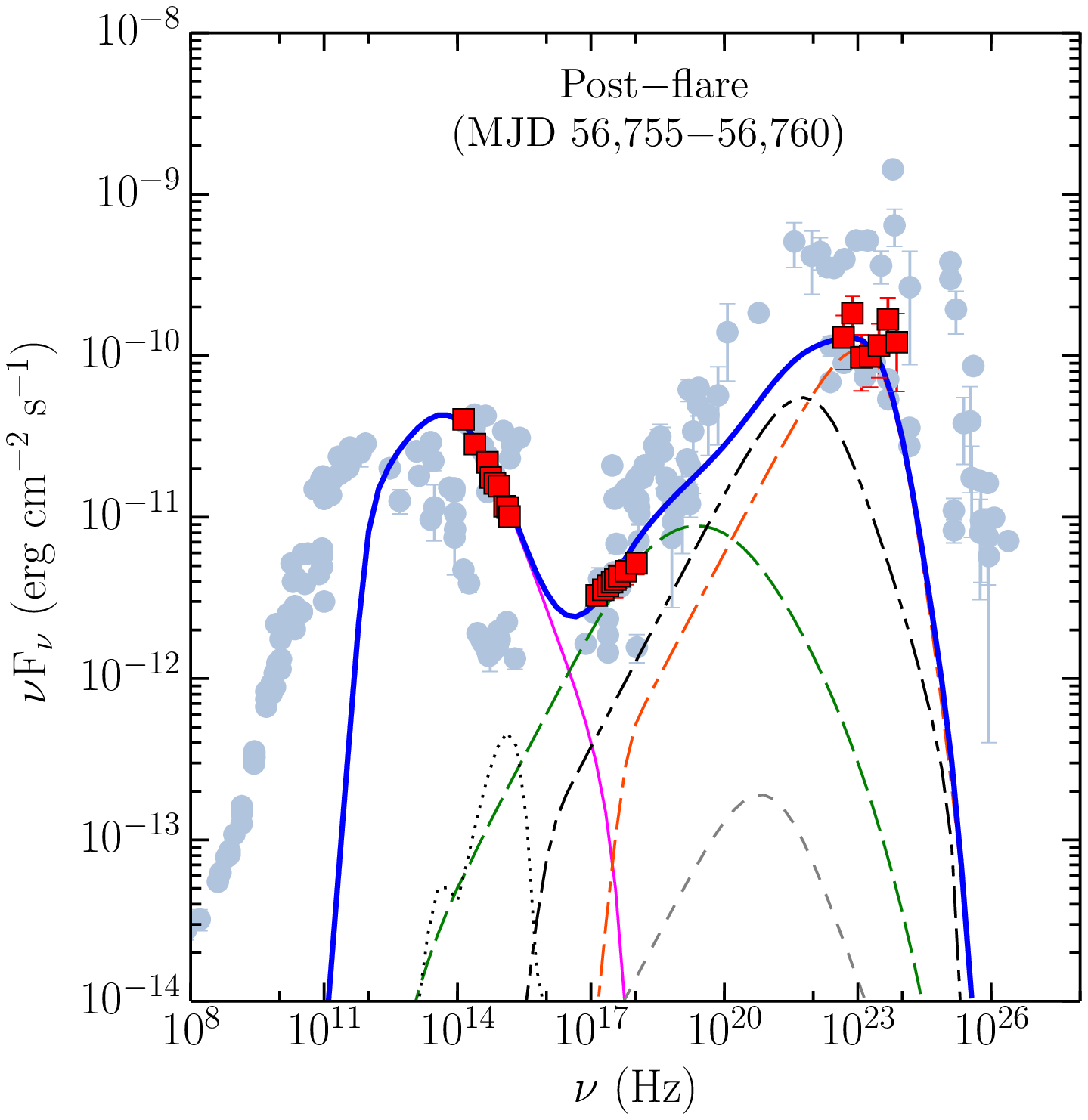}
     }
\caption{Spectral energy distributions of 3C 279 during low and high activity states. Simultaneous data from SMARTS, {\it Swift} and \fermi-LAT are shown with red squares whereas light blue circles belongs to archival observations. Black dotted line represents thermal contributions from the torus, accretion disk, and X-ray corona (not in the plots). Pink thin solid and green dashed lines correspond to synchrotron and SSC emission respectively. Grey dashed, red dash-dot, and black dash-dot-dot lines represent EC-disk, EC-BLR, and EC-torus components respectively. Blue thick solid line is the sum of all the radiative mechanisms.}\label{fig:sed_fit}
\end{center}

\end{figure*}

\begin{figure*}
\hbox{
      \includegraphics[width=\columnwidth]{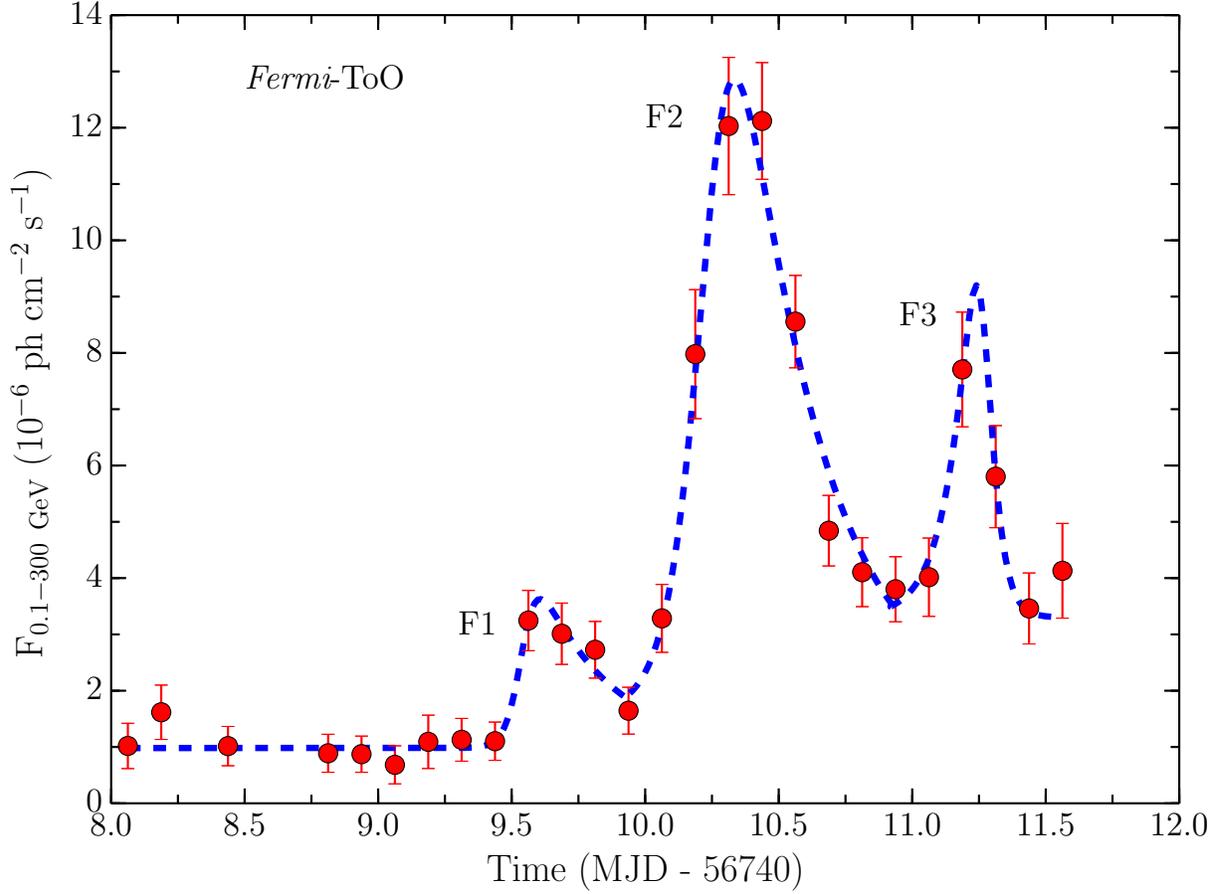}
     }
\caption{Three hour binned $\gamma$-ray light curve of 3C 279 using the data taken in the pointed mode of the \fermi-LAT ToO observations. Only those bins when TS $>$ 9, are used for fitting. F1, F2, and F3 correspond to three flares for which fitting is performed. Dashed blue line represent the best-fit temporal profiles assuming an exponential rise and fall.}\label{fig:flare_profile}
\end{figure*}

A recent study on the multi-wavelength behavior of 3C 279 \citep{2012ApJ...754..114H} reported a significant correlation between optical and $\gamma$-rays for the period 2008 to 2010. However, they have not found a correlation between variations in X-ray and $\gamma$-ray bands. Another study of 3C 279 in 2011 by \citet{2014A&A...567A..41A} has led to the conclusion that X-ray and $\gamma$-ray flux variations are correlated whereas no significant correlation is seen between optical and $\gamma$-rays. Such inconsistent patterns of correlations are already seen in the long term variability studies of 3C 279 \citep[e.g.,][]{2008ApJ...689...79C}. For the flare under consideration in this work, visual inspection of the multi-frequency light curves (Figure~\ref{fig:mw_lc}) reveals the enhancement of the fluxes in all the wavebands. This suggests that a single emission region as well as the same electron population are responsible for the flaring emission. However, generation of high temporal resolution $\gamma$-ray light curves during the flaring period reveals the presence of multiple shorter timescale flaring events. During the Flare 1 phase, two major flares are observed in twelve hours binned light curve (one before MJD 56,746 and other after twelve hours). Interestingly, the former seems to be resolved in six and three hour bins, whereas the latter is non-resolvable down to three hours. Further, to understand the nature of the flares that occurred during the Fermi ToO observations (where the data are having better S/N than other periods), we use the following equation to fit the flare profiles \citep[see e.g.,][]{2010ApJ...722..520A}
\begin{equation}\label{eq:flare_profile}
F(t) = F_c + F_p \left(e^{\frac{t_p - t}{T_r}} + e^{\frac{t-t_p}{T_f}}\right)^{-1}
\end{equation}
where $F_c$ represents an assumed constant level underlying the flare, $F_p$ measures the amplitude of the flare, $t_p$ describes an approximate time of peak, and $T_r$ and $T_f$ are the rise and fall time. The results of the fitting of three flares (F1, F2, and F3) are shown in Figure~\ref{fig:flare_profile} and the associated parameters are given in Table~\ref{tab:flare_profile}. Barring the third flare where we could not get the reliable parameters, remaining two flares display a clear trend of fast rise and slow decay that can be interpreted as a result of particle acceleration mechanism. A fast rise of the flare could be attributed to the higher rate of acceleration, probably at a shock front, and the slow decay can be associated with the weakening of the shock.

From SED modeling, we find the location of the emission region at the outer edge of the BLR where the total energy density of the external soft photons are provided by the BLR clouds and the dusty torus in roughly equal fractions (see Figure~\ref{fig:sed_fit}). The primary mechanism for the production of $\gamma$-rays, thus, would be the IC scattering of BLR and torus photons. Accordingly, the cooling timescale for the electrons responsible for the emission of $\gamma$-rays ($\epsilon_{\gamma}$ = 1 GeV), measured in the observer’s frame, would be \citep[e.g.,][]{2013ApJ...766L..11S}
\begin{equation}\label{eq:cooling_time}
\tau_{\rm rad} \simeq (3m_{e}c/4\sigma_{\rm T}u'_{\rm total}) \times [\epsilon_0(1+z)/\epsilon_{\gamma}]^{0.5},
\end{equation}
i.e. $\sim$7 minutes. Here, $\epsilon_0$ is the characteristic energy of the seed photons (10.2 eV for BLR and 0.27 eV for torus photons) and $u'_{\rm total}$ is the total seed photon energy density in the comoving frame. The obtained cooling time is significantly shorter than the observed shortest flux halving time and the decay time of the flares (Table~\ref{tab:shortest_gamma_var}, \ref{tab:flare_profile}). This suggests that the flare timescale is governed by the processes other than radiative losses, probably associated with particle acceleration or jet dynamics \citep{2009ApJ...692.1374B,2014MNRAS.442..131K}. Alternatively, the flare timescale can also be associated with the geometry and presence of sub-structures in the emitting region \citep[see e.g.,][]{2001ApJ...563..569T}.

The maximum $\gamma$-ray flux during the period of high activity is found to be (1.21 $\pm$ 0.10) $\times$ 10$^{-5}$ \phflux~which corresponds to an isotropic $\gamma$-ray luminosity of 1.2 $\times$ 10$^{49}$ \lum. Adopting the bulk Lorentz factor $\Gamma$ = 19 (obtained from the SED modeling of the Flare 2),
the total power emitted in the $\gamma$-ray band, in the proper frame of the jet, would be $L_{\gamma, em} \simeq L_{\gamma}/2\Gamma^{2} \simeq$ 1.7 $\times$ 10$^{46}$ \lum. This is a good fraction of the kinetic jet power ($\sim$ 23\%; $P_{\rm j,kin}$ = 7.2 $\times$ 10$^{46}$ \lum). This implies that the jet becomes radiatively efficient and a significant amount of the kinetic jet power is released in the form of high energy $\gamma$-ray radiation. Comparing with the Eddington luminosity ($L_{\rm Edd}$, for a black hole mass of 3 $\times$ 10$^{8}$ \Msun), $L_{\gamma, em}$ is about $\sim$ 45\% of $L_{\rm Edd}$, thus supporting high radiative efficiency of the jet. Moreover, $L_{\gamma, em}$ is found to be 1.7 times larger than the total available accretion power ($L_{\rm acc} \simeq L_{\rm disk}/\eta_{\rm disk} \simeq 1 \times 10^{46}$ erg s$^{-1}$; assuming radiative efficiency $\eta_{\rm disk}$ = 10\%). Recently, it has been established by \citet{2014Natur.515..376G} that, in blazars, the radiative jet power (to which the $L_{\gamma, em}$ is a good proxy) is of the same order of the accretion disk luminosity. The parameters obtained here, thus, indicate for the extremely efficient conversion of the accretion power and/or kinetic jet power to the jet $\gamma$-ray luminosity. Similar results have been found by \citet{2013ApJ...766L..11S} for the GeV outburst of FSRQ PKS 1510$-$089. However, it should be noted that such events are short-lived only. This is due to the fact that the fraction of time in which the source is in a flaring state is about 1\% \citep{2010MNRAS.405L..94T}.

Comparing the SEDs corresponding to low and flaring activity states, we find that the flux has substantially increased across the electromagnetic spectrum. However, considering the SEDs during the three sub-phases of the flare, the major flux enhancement is observed in the $\gamma$-ray band, whereas, relatively lesser degree of flux variations are seen in the optical$-$UV and X-ray bands. This is also evident in the light curves shown in Figure~\ref{fig:mw_lc}. These changes are explained primarily by varying the bulk Lorentz factor, magnetic field, and particle energy density (Table~\ref{tab:sed}). In addition to these, there are minor changes in other parameters such as the location of the emission region (and hence the emission region size), and spectral indices of the electron energy distributions. These changes are required to explain the variations in the X-ray spectra, in particular the post$-$flare X-ray spectrum, which is significantly softer than the other two sub-phases. These modifications, though minor, lead to large variations in the total jet power, because $\gamma'_{\rm b}$ and $p$ decides the total amount of electrons present in the emission region and thus the total number of cold protons (assuming both of them to have equal number densities). Interestingly, the maximum jet power is found during the post$-$flare rather than the peak of the $\gamma$-ray flux.

During the Flare 2, a significant curvature in the $\gamma$-ray spectrum is noticed (see Figure~\ref{fig:gamma_spec} and Table~\ref{tab:gamma_spec}). Similar feature is also noticed in the $\gamma$-ray spectrum of FSRQ 3C 454.3 during its giant outburst in 2010 November \citep{2011ApJ...733L..26A}. \citet{2010ApJ...717L.118P} have proposed a possible explanation of such curvature as due to the attenuation of $\gamma$-rays by photon-photon pair production on He {\sc ii} Lyman recombination lines within the BLR. Recently, \citet{2013ApJ...771L...4C} have explained the origin of curved $\gamma$-ray spectrum due to Klein-Nishina (KN) effect and log-parabolic electron distribution. We reproduce the observed curvature in the $\gamma$-ray spectrum by KN mechanism and with broken power law electron distribution. In addition, we also adjust the BLR and torus energy densities in such a manner that the observed curvature can be explained by the superposition of these external photon fields \citep[see a similar approach followed in][]{2013ApJ...771L...4C}. Since in our model, the radiation energy densities are a function of the distance from the central black hole, reproduction of the curvature in $\gamma$-ray window of the SED can possibly hint the location of the emission region. Similar approach is followed by \citet{2014ApJ...782...82D} to explain the $\gamma$-ray spectra of 3C 279, though they also consider equi-partition between various photon fields, particle energy density, and magnetic field. Interestingly, both the studies \citep[i.e.,][]{2013ApJ...771L...4C,2014ApJ...782...82D} concluded the location of the emission region to be at the outer edge of the BLR, akin to our findings. Moreover, a curvature in the $\gamma$-ray spectrum also suggests significant absorption of the VHE photons by the BLR radiation field (in the context of the above mentioned discussion). Since we find a curvature during the Flare 2, the probability of detecting VHE emission from 3C 279 during this period should be quite low. In fact, \fermi-LAT did not detect any VHE photon from 3C 279 during the entire flaring period (see Section~\ref{subsec:highest_gamma_photon}). Additionally, preliminary results of the Very Energetic Radiation Imaging Telescope Array System (VERITAS) observations taken during the Flare 2 also indicates the non-detection of VHE events from 3C 279 \citep{2014HEAD...1410611E}\footnote{http://files.aas.org/head14/106-11\_Manel\_Errando.pdf}. The cut-off at GeV frequencies as hinted by the curved LAT spectrum, upper limits in the VERITAS observations, and the measured short timescale variability, therefore suggests that the emission region cannot be far outside from the BLR.

The optical polarization monitoring from the Steward observatory indicates the anti-correlated behavior of the optical polarization with respect to the $\gamma$-ray flux (Figure~\ref{fig:mw_lc}). During the Flare 1 period, an enhancement in the $\gamma$-ray flux can be seen but the optical polarization behaves opposite. At the same time, variation of the polarization angle follows similar trend as seen in the $\gamma$-ray flux. Similar behavior from 3C 279 was earlier observed during its 2009 flare \citep{2010Natur.463..919A}. Interestingly, though the $\gamma$-ray flux level during the recent outburst is much higher than that seen during the 2009 flare, the change in the optical polarization and polarization angle is relatively smaller. A possible reason could be that the $\gamma$-ray flare may be associated with the change in the optical polarization, however, not to a single coherent event, but due to the superposition of multiple shorter duration events. In such a scenario, though the $\gamma$-ray flux increases, the overall polarization gets averaged out resulting in lesser degree of change of polarization. This is supported by the short duration flares seen in the three hour binned $\gamma$-ray light curve. Moreover, as discussed below, we explain the flare due to bulk acceleration of the jet which may not be directly related to the optical polarization and hence though there is significant change in the flux level, it is not reflected in the polarization observations. Unfortunately, we do not have polarimetric observations for the period of Flare 2 and thus it is not possible to predict the polarization behavior during the peak of the $\gamma$-ray activity.

\begin{figure*}
\hbox{
      \includegraphics[width=\columnwidth]{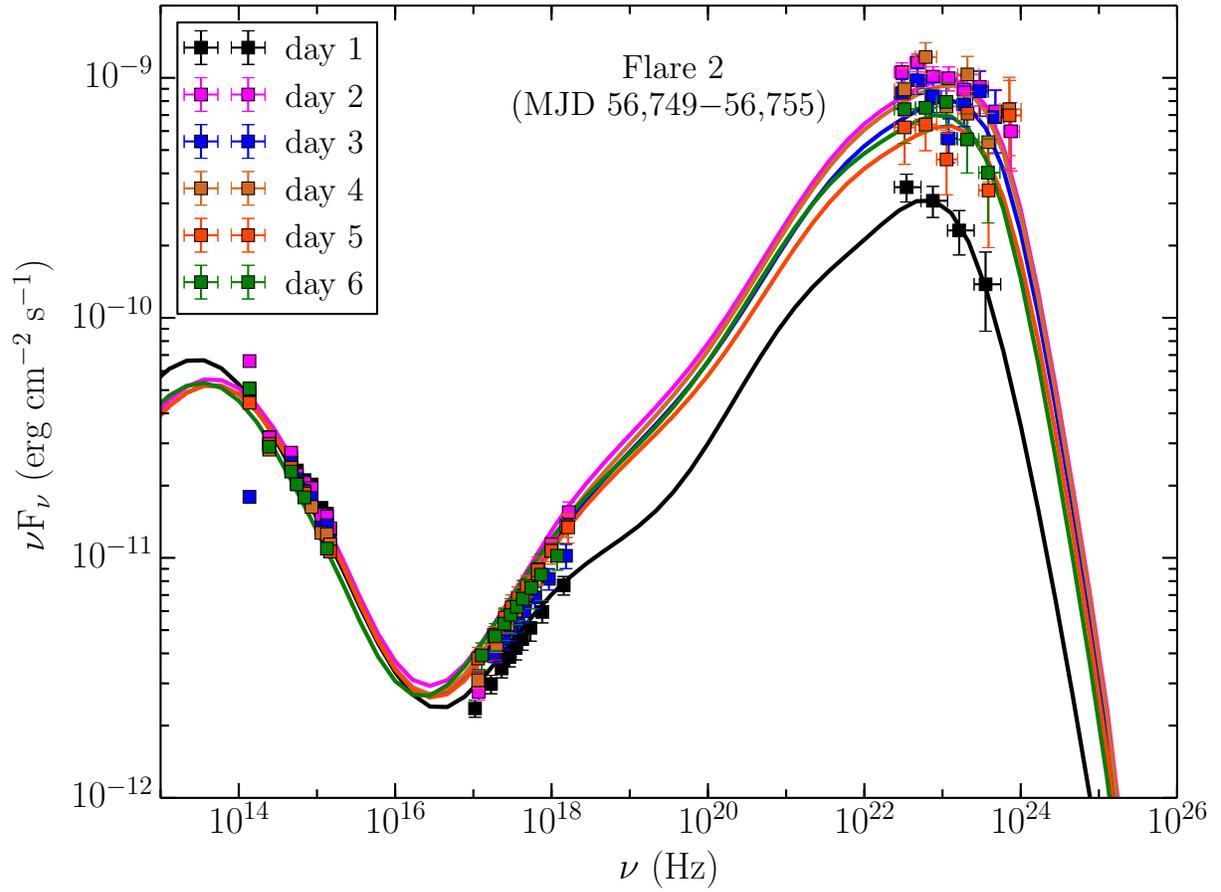}
     }
\caption{One day averaged SEDs of 3C 279 covering the main flaring period.}\label{fig:sed_all}
\end{figure*}

\begin{figure*}
\hbox{
      \includegraphics[width=\columnwidth]{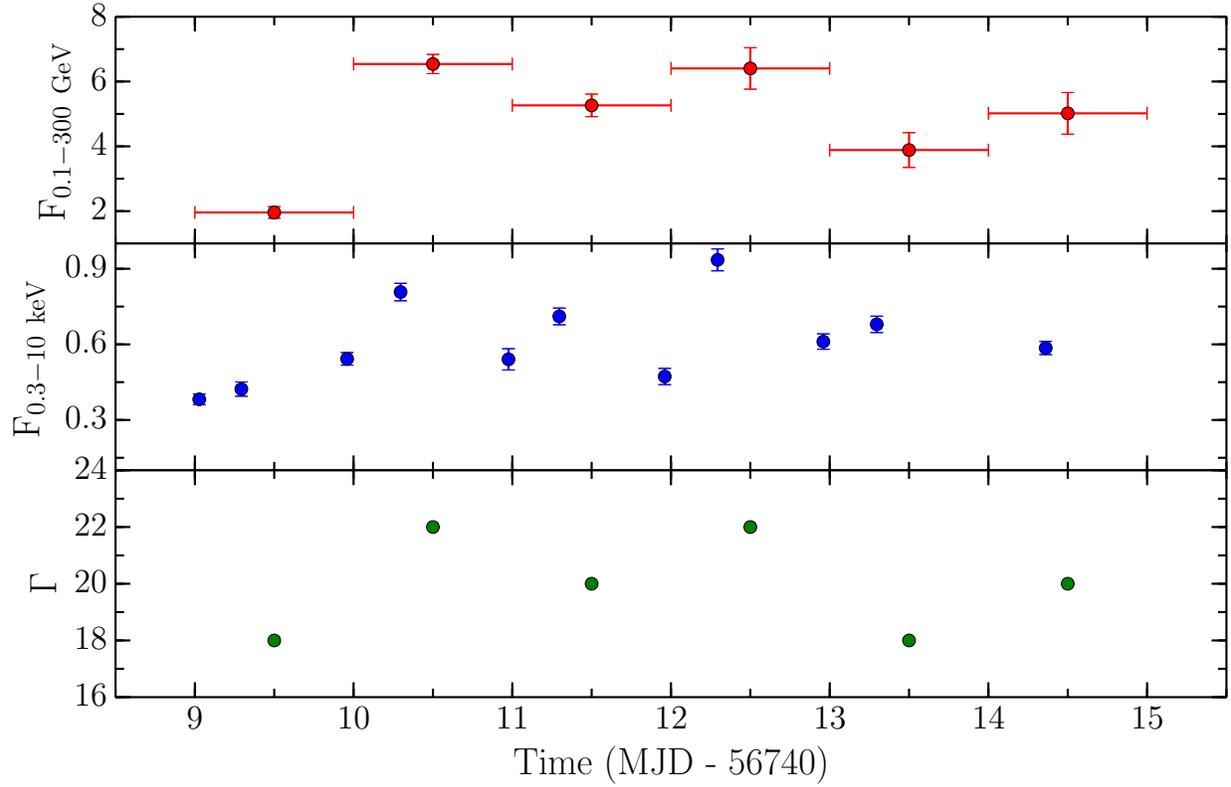}
     }
\caption{Variation of bulk Lorentz factor $\Gamma$ as a function of time (bottom panel). For comparison, the variation of X-ray and $\gamma$-ray fluxes are also shown (top two panels). Units are same as in Figure~\ref{fig:mw_lc}. Horizontal errorbar represent the time ranges.}\label{fig:flux_lorentz}
\end{figure*}

{\it Swift} and SMARTS have monitored 3C 279 almost every day during the Flare 2 period. This has enabled us to study the time evolution of the flare by generating the SEDs using finer time bins. In Figure~\ref{fig:sed_all}, we show the one day averaged SEDs covering the duration of the Flare 2 and the corresponding modeling parameters are given in Table~\ref{tab:sed_1day}. In comparison to the optical and X-rays, a greater degree of enhancement in the $\gamma$-ray flux can be noticed. Since the optical and X-rays are due to synchrotron and SSC processes respectively and, $\gamma$-rays is due to EC mechanism, this difference could hint the possible change in the source parameters responsible for the flare. The synchrotron emissivity in the comoving frame can be approximated as \citep[e.g.,][]{1991pav..book.....S}
\begin{equation}
\label{eq:appaemiss}
j'_{\rm syn}(\epsilon')\approx\frac{\sigma_{\rm T}cB^2}{48 \pi^2}\epsilon_{\rm L}^{-\frac{3}{2}}
  N'\left(\sqrt{\frac{\epsilon'}{\epsilon_{\rm L}}}\right)\epsilon^{\prime{\frac{1}{2}}}
\end{equation}
where $\epsilon_{\rm L}$ = ($h\nu_{\rm L}/mc^2$) is the quanta (dimensionless) corresponding to the Larmor frequency. On the other hand, the EC emissivity can be approximated as \citep{1995ApJ...446L..63D,2012MNRAS.419.1660S}
\begin{equation}
\label{eq:ecemiss}
j'_{\rm ec}(\epsilon')\approx\frac{c\sigma_{\rm T}u^\star}{8\pi\epsilon^\star}
\left(\frac{\Gamma \epsilon'}{\epsilon^\star}\right)^{\frac{1}{2}} 
N'\left[\left(\frac{\epsilon'}{\Gamma\epsilon^\star}\right)^{\frac{1}{2}}\right]
\end{equation}
where starred quantities are in the AGN frame. Comparing Equation~\ref{eq:appaemiss} and \ref{eq:ecemiss} we find that the excess in EC emissivity can be obtained by increasing the jet bulk Lorentz factor without altering the synchrotron emissivity. Increase in $\Gamma$ will also result in further boosting of observed synchrotron, SSC, and EC fluxes in addition to the increase in emissivity of the EC. To illustrate this, in Figure~\ref{fig:flux_lorentz} we show the variation of $\Gamma$, obtained from the SED modeling, along with the variation of X-ray and $\gamma$-ray fluxes. Clearly, the pattern of variability seen in X-rays and $\gamma$-rays is similar to that obtained in $\Gamma$.

Recently, \citet{2014A&A...567A..41A} have reported the multi-wavelength study of 3C 279 covering a low and a high activity period in 2011. In both the activity states, the source was monitored by MAGIC telescopes. They could not ascertain the location of the emission region in the low activity state because the SED and the relevant modeling parameters are found to be satisfactorily explained by both inside the BLR and outside the BLR scenario. However, absence of the simultaneous data points covering NIR to UV frequencies hampers their interpretation (see their Figure 8). This is because the slope of the high energy synchrotron spectrum constrains the shape of the falling IC spectrum which lies in the LAT energy range (assuming the same electron population is responsible for the emission at both regimes). The lack of the optical$-$UV data points, thus, could lead to the degeneracy in reproducing the SEDs. In contrast, the availability of the contemporaneous data from NIR to UV in the low activity state SED modeled by us, not only constrains the slope of the falling synchrotron spectrum but also the accretion disk radiation. It can be seen in the low activity state SED in Figure~\ref{fig:sed_fit} that either EC-BLR or EC-torus alone cannot explain the observed $\gamma$-ray spectrum and hence a combination of both is required. This constrains the location of the emission region as, in our model, the radiation energy densities are a function of dissipation distance from the central black hole. Further, \citet{2014A&A...567A..41A} have fitted the SED of a high activity state of 3C 279 with a two-zone leptonic model. This choice was driven by the observed correlated X-ray and $\gamma$-ray variations as well as by the lack of correlations seen between the optical and other wavebands. In this two-zone model, the X-ray to $\gamma$-ray emitting region lies inside the BLR (so as to explain the LAT spectrum and MAGIC upper limits) whereas low energy emitting region lies outside the BLR. During the high activity state studied in our work, we find enhancement in fluxes at all the wavelengths which, unlike the 2011 flare, supports the single zone origin of the radiations. However, similar to them, the shape of the LAT spectrum and non-detection by VERITAS suggest that the BLR has significant impact on the observed $\gamma$-ray spectrum and thus the emission region cannot be far out from it.

\citet{2012ApJ...754..114H} have studied 3C 279 using an extensive broadband data set covering the first two years of \fermi operation. They report the presence of double synchrotron peaks at mid to far IR frequencies and a delay of about 10 days between optical and $\gamma$-rays as found from cross-correlation studies. Interestingly, they argue that X-rays do not correlate with optical and $\gamma$-ray fluxes during the flaring states and hence X-ray data are not accounted for their SED modeling. In comparison to that, we find similar behavior of the fluxes at different wavelengths. In the modeling performed by the \citet{2012ApJ...754..114H}, the location of the emission region is constrained on the basis of the observed change of optical polarization and associated rotation of the electric vector polarization angle (EVPA), which was accompanied by the $\gamma$-ray flare \citep[see also][]{2010Natur.463..919A}. Though the flux amplitude is much higher during the 2014 $\gamma$-ray flare, the corresponding rotation of the EVPA as well as change in the optical polarization are found to be much lesser than that obtained during the 2009 flare. However, we stress that we do not have polarization data during the main flaring event and thus a strong claim regarding the variability of optical polarization associated with the $\gamma$-ray flaring event cannot be made. There are few other differences such as they use a comparatively long variability time scale ($\approx$ 2 weeks) and double broken power law electron energy distribution. Their modeled magnetic field is also relatively lower compared to the one obtained by us.

In the model by \citet{2012MNRAS.419.1660S}, where the authors discuss the 2006 flare of 3C 279, the emission region is assumed to be far out from the BLR, to avoid severe attenuation of VHE $\gamma$-rays by the BLR Lyman-$\alpha$ line emission. Their model overpredicts the $\gamma$-ray flux at MeV energies by about an order higher than that ever observed from 3C 279. Since there are no MeV-GeV observations available at the time of flare, this possibility cannot be ruled out. However, as discussed by \citet{2014ApJ...782...82D}, fitting the VHE data (along with the LAT spectrum) with single zone leptonic emission models will result in the parameters well out of equipartition. Alternatives to avoid such issues could be the use of multi-zone emission modeling or inclusion of hadronic radiation scenario \citep[see e.g.,][]{2009ApJ...703.1168B}. Since during the 2014 flare, there is no detection of VHE events and the $\gamma$-ray spectrum is curved, the parameters obtained by us (under near-equipartition condition) using a single zone leptonic emission model seems to be robust. 

\section{Summary}\label{sec:summary}
In this paper, a detailed study of the brightest $\gamma$-ray flare observed from 3C 279 in 2014 March$-$April is presented. Below we summarize our main findings.
\begin{enumerate}
\item In the energy range of 0.1$-$300 GeV, the maximum $\gamma$-ray flux of (1.21 $\pm$ 0.10) $\times$ 10$^{-5}$ \phflux~is observed on MJD 56,750. This is the highest $\gamma$-ray flux detected from 3C 279, since the launch of \fermi~satellite.
\item The shortest $\gamma$-ray flux doubling time measured is 1.19 $\pm$ 0.36 hours, on MJD 56,746.
\item A significant curvature is noticed in the $\gamma$-ray spectrum during the Flare 2. This suggests the low probability of detecting VHE events and is confirmed by the non-detection by VERITAS.
\item During the flare, the jet becomes radiatively efficient and a good fraction of kinetic power gets converted to high energy $\gamma$-ray radiation.
\item During the Flare 1 phase, variation in the optical polarization and the rotation of EVPA are smaller compared to 2009 flare. However, we do not have polarization measurement during the Flare 2 and thus predicting the polarization behavior during the peak of the $\gamma$-ray emission is not possible.
\item A simple one zone leptonic emission model satisfactorily explains the observed SEDs with increase in the bulk Lorentz factor as a major cause of the flare.
\item Observations such as the presence of a significant curvature in the $\gamma$-ray spectrum, short time scale of variability, and lack of VHE $\gamma$-rays suggest the location of the emission region to be at the outer edge of the BLR where both BLR and torus energy densities are contributing to the observed $\gamma$-ray spectrum.
\end{enumerate}
\acknowledgments
We thank the referee for a constructive report that helped to improve the manuscript. VSP is grateful to Gang Cao for useful discussions and suggestions. This research has made use of data, software and/or web tools obtained from NASA’s High Energy Astrophysics Science Archive Research Center (HEASARC), a service of Goddard Space Flight Center and the Smithsonian Astrophysical Observatory. Part of this work is based on archival data, software, or online services provided by the ASI Science Data Center (ASDC). This research has made use of the XRT Data Analysis Software (XRTDAS) developed under the responsibility of the ASDC, Italy. Data from the Steward Observatory spectropolarimetric monitoring project were used. This program is supported by Fermi Guest Investigator grants NNX08AW56G, NNX09AU10G, and NNX12AO93G. This paper has made use of up-to-date SMARTS optical/near-infrared light curves that are available at www.astro.yale.edu/smarts/glast/home.php. Use of {\it Hydra} cluster at the Indian Institute of Astrophysics is acknowledged.

{\it Facilities:} \facility{\fermi}, \facility{{\it Swift}}, \facility{SMARTS}, \facility{Steward}.
\bibliography{Master}
\bibliographystyle{apj}

\begin{table}
\center
\caption{Fractional root mean square variability amplitude ($F_{\rm var}$) values for different energy bands, calculated for the light curves shown in Figure~\ref{fig:mw_lc}. Time periods during which observations were made and total number of observations in that period are also given.}\label{tab:f_var}
 \begin{tabular}{lccc}
\hline \hline
Energy band & $F_{\rm var}$ & Time period & Number of observations\\
\hline
K (SMARTS)          & 0.227 $\pm$ 0.001 & 56,742$-$56,760 & 15 \\
J (SMARTS)          & 0.067 $\pm$ 0.002 & 56,742$-$56,760 & 16 \\
V (Steward)         & 0.062 $\pm$ 0.002 & 56,741$-$56,749 & 20 \\
V (SMARTS)          & 0.074 $\pm$ 0.001 & 56,742$-$56,760 & 16 \\
V (UVOT)            & 0.083 $\pm$ 0.020 & 56,745$-$56,760 &  6 \\
B (SMARTS)          & 0.089 $\pm$ 0.002 & 56,742$-$56,760 & 16 \\
B (UVOT)            & 0.072 $\pm$ 0.016 & 56,745$-$56,760 &  7 \\
U (UVOT)            & 0.081 $\pm$ 0.016 & 56,745$-$56,760 &  8 \\
UVW1                & 0.097 $\pm$ 0.020 & 56,745$-$56,760 &  9 \\
UVM2                & 0.119 $\pm$ 0.021 & 56,745$-$56,760 &  7 \\
UVW2                & 0.096 $\pm$ 0.018 & 56,745$-$56,760 & 10 \\
X-ray (0.3$-$10 keV)  & 0.347 $\pm$ 0.014 & 56,743$-$56,760 & 15 \\
$\gamma$-ray (0.1$-$300 GeV) & 0.739 $\pm$ 0.034 & 56,741$-$56,760 & 19 \\
\hline
\end{tabular}
\end{table}

\begin{table}
\caption{Summary of the Search for the Shortest Time Scale of Variability Using Three Hours Binned $\gamma$-ray Light Curve.}\label{tab:shortest_gamma_var}
 \begin{tabular}{ccccccc}
\hline \hline
$t$ & $t_0$ & $F(t)$ &  $F(t_0)$ & $|\tau|$ & Signif. & R/D \\
\hline
56746.4376 & 56746.5626 &  1.91 $\pm$ 0.94 & 10.95 $\pm$ 1.92 & 1.19 $\pm$ 0.36 & 4.225 & R \\
56746.5626 & 56746.6876 & 10.95 $\pm$ 1.92 & 2.94 $\pm$ 1.12  & 1.58 $\pm$ 0.51 & 3.599 & D \\
56749.4376 & 56749.5626 & 1.03 $\pm$ 0.37  & 3.33 $\pm$ 0.57  & 1.78 $\pm$ 0.60 & 3.400 & R \\
56750.0626 & 56750.1876 & 3.29 $\pm$ 0.65  & 7.57 $\pm$ 1.16  & 2.49 $\pm$ 0.75 & 3.219 & R \\
56750.5626 & 56750.6876 & 8.34 $\pm$ 0.85  & 4.50 $\pm$ 6.26  & 3.37 $\pm$ 0.94 & 3.636 & D \\
\hline
\end{tabular}
\tablecomments{Times $t$ and $t_0$ are in MJD; fluxes are in units of 10$^{-6}$ \phflux; the absolute values of the observed characteristic time scale $|\tau|$ are in hours; significance of flux differences is in $\sigma$; R and D denote the rise or a decay time.}
\end{table}

\begin{table}
\caption{Results of the Model Fitting to the $\gamma$-ray Spectra of 3C 279, obtained for different time periods. Col.[1]: period of observation (MJD); Col.[2]: activity state; Col.[3]: model used (PL:
power law, LP: logParabola); Col.[4]: integrated $\gamma$-ray flux (0.1$-$300 GeV), in units of 10$^{-6}$
\phflux; Col.[5] and [6]: spectral parameters (see definitions in the text); Col.[7]: test statistic; Col.[8]: $TS_{\rm curve}$ .}\label{tab:gamma_spec}
\begin{center}
\begin{tabular}{cccccccc}
\hline\hline
Period & Activity & Model & $F_{0.1-300~{\rm GeV}}$ & $\Gamma_{0.1-300~{\rm GeV}}/\alpha$ & $\beta$ & TS & $TS_{\rm curve}$\\
~[1] & [2] & [3] & [4] & [5] & [6] & [7] & [8] \\
\hline
55,300$-$55,400 & Low activity & PL & 0.16 $\pm$ 0.02 & 2.48 $\pm$ 0.08 &                 & 315.61  & \\
                &              & LP & 0.15 $\pm$ 0.02 & 2.42 $\pm$ 0.13 & 0.04 $\pm$ 0.06 & 315.50  & 0.01 \\
56,741$-$56,749 & Flare 1      & PL & 1.64 $\pm$ 0.10 & 2.15 $\pm$ 0.05 &                 & 1276.40 & \\
                &              & LP & 1.53 $\pm$ 0.10 & 1.96 $\pm$ 0.09 & 0.10 $\pm$ 0.04 & 1286.53 & 7.74 \\
56,749$-$56,755 & Flare 2      & PL & 4.47 $\pm$ 0.15 & 2.23 $\pm$ 0.03 &                 & 4445.66 & \\
                &              & LP & 4.24 $\pm$ 0.15 & 2.05 $\pm$ 0.05 & 0.13 $\pm$ 0.03 & 4463.94 & 29.96 \\
56,755$-$56,760 & Post$-$flare & PL & 0.95 $\pm$ 0.16 & 2.20 $\pm$ 0.11 &                 & 204.39  & \\
                &              & LP & 0.83 $\pm$ 0.17 & 1.93 $\pm$ 0.23 & 0.13 $\pm$ 0.10 & 206.04  & 2.31 \\
\hline
\end{tabular}
\end{center}
\end{table}

\begin{table*}
\begin{center}
{
\small
\caption{Summary of the SED analysis. \fermi-LAT analysis results are given in Table~\ref{tab:gamma_spec}.}\label{tab:sed_flux}
\begin{tabular}{ccccccc}
\tableline\tableline
 & & & {\it Swift}-XRT  & & &\\
 Activity state & Exp.\tablenotemark{a} & $\Gamma_{0.3-10~{\rm keV}}$\tablenotemark{b} & $F_{0.3-10~{\rm keV}}$\tablenotemark{c} & Norm.\tablenotemark{d} & Stat.\tablenotemark{e} & \\
 \tableline
 Low activity & 5.97  & 1.62$^{+0.06}_{-0.06}$   & 1.01$^{+0.06}_{-0.06}$ & 1.36$^{+0.07}_{-0.07}$  & 69.10/59 & \\
 Flare 1  & 2.49  & 1.53$^{+0.08}_{-0.07}$   & 2.01$^{+0.15}_{-0.14}$ & 2.48$^{+0.15}_{-0.15}$  & 44.68/47 & \\
 Flare 2  & 10.16 & 1.47$^{+0.03}_{-0.03}$   & 2.54$^{+0.09}_{-0.08}$ & 2.93$^{+0.08}_{-0.08}$  & 172.50/179&\\
 Post$-$flare & 0.93  & 1.77$^{+0.15}_{-0.15}$   & 1.01$^{+0.20}_{-0.17}$ & 2.35$^{+0.25}_{-0.25}$  & 16.14/14 & \\
 \tableline
 & & & {\it Swift}-UVOT &  & & \\
 Activity state & V\tablenotemark{f} & B\tablenotemark{f} & U\tablenotemark{f} & UVW1\tablenotemark{f} & UVM2\tablenotemark{f} & UVW2\tablenotemark{f} \\
 \tableline
Low activity & 0.13 $\pm$ 0.01 & 0.11 $\pm$ 0.01 & 0.09 $\pm$ 0.01 & 0.10 $\pm$ 0.00 & 0.09 $\pm$ 0.00 & 0.09 $\pm$ 0.00 \\
Flare 1 & 1.95 $\pm$ 0.09 & 1.93 $\pm$ 0.05 & 1.75 $\pm$ 0.05 & 1.31 $\pm$ 0.05 & 1.19 $\pm$ 0.07 & 1.11 $\pm$ 0.04 \\
Flare 2 & 2.18 $\pm$ 0.05 & 1.99 $\pm$ 0.04 & 1.80 $\pm$ 0.03 & 1.40 $\pm$ 0.03 & 1.35 $\pm$ 0.03 & 1.24 $\pm$ 0.02 \\
Post$-$flare & 1.76 $\pm$ 0.09 & 1.62 $\pm$ 0.07 & 1.56 $\pm$ 0.07 & 1.17 $\pm$ 0.07 & 1.15 $\pm$ 0.06 & 1.01 $\pm$ 0.06 \\
\tableline
 & & & SMARTS &  & & \\
 Activity state &  & R\tablenotemark{g} & J\tablenotemark{g} & K\tablenotemark{g} &  &  \\
 \tableline
Low activity &  & 0.16 $\pm$ 0.00 & 0.23 $\pm$ 0.00 & 0.49 $\pm$ 0.00 &  &  \\
Flare 1      &  & 2.41 $\pm$ 0.00 & 2.84 $\pm$ 0.00 & 4.35 $\pm$ 0.00 &  &  \\
Flare 2      &  & 2.49 $\pm$ 0.00 & 3.01 $\pm$ 0.01 & 4.62 $\pm$ 0.00 &  &  \\
Post$-$flare &  & 2.19 $\pm$ 0.01 & 2.83 $\pm$ 0.01 & 4.04 $\pm$ 0.01 &  &  \\
 \tableline
\end{tabular}
\tablenotetext{1}{Net exposure in kiloseconds.}
\tablenotetext{2}{Photon index of the absorbed power law model.}
\tablenotetext{3}{Observed flux in units of 10$^{-11}$ erg cm$^{-2}$ s$^{-1}$, in 0.3$-$10 keV energy band.}
\tablenotetext{4}{Normalization at 1 keV in 10$^{-3}$ \phflux~keV$^{-1}$.}
\tablenotetext{5}{Statistical parameters: $\chi^2$/dof.}
\tablenotetext{6}{Average flux in {\it Swift} V, B, U, W1, M2, and W2 bands, in units of 10$^{-11}$ erg cm$^{-2}$ s$^{-1}$.}
\tablenotetext{7}{Average flux in SMARTS R, J, and K bands, in units of 10$^{-11}$ erg cm$^{-2}$ s$^{-1}$.}
}
\end{center}
\end{table*}

\begin{table*}
{\scriptsize
\begin{center}
\caption{Summary of the parameters used/derived from the modeling of the SEDs in Figure~\ref{fig:sed_fit}. Viewing angle is taken as 3$^{\circ}$ and the characteristic temperature of the torus as 800 K. For a disk luminosity of 1 $\times$ 10$^{45}$ erg s$^{-1}$ and black hole mass of 3 $\times$ 10$^8$ \Msun, the size of the BLR is 0.03 parsec (1029 $R_{\rm Sch}$).}\label{tab:sed}
\begin{tabular}{lcccc}
\tableline
\tableline
Parameter                                  &  Low activity& Flare 1 & Flare 2 & Post$-$flare  \\
\tableline
$p$\tablenotemark{a}                            & 1.65         & 1.65        & 1.65    & 1.95     \\
$q$\tablenotemark{b}                            & 5.2          & 4.5         & 4.5     & 4.5     \\
$B$\tablenotemark{c}                            & 1.2          & 2.6         & 2.0     & 3.0     \\
$U'_{\rm e}$\tablenotemark{d}                   & 0.15         & 0.16        & 0.18    & 0.16     \\
$\Gamma$\tablenotemark{e}                       & 10           & 13          & 19      & 14     \\
$\gamma'_{\rm b}$\tablenotemark{f}              & 647          & 689         & 758     & 687     \\
$\gamma'_{max}$\tablenotemark{g}                & 1e4          & 3e4         & 5e4     & 3e4     \\
$R_{\rm diss}$\tablenotemark{h}                 & 0.045 (1570) & 0.034 (1190) & 0.035 (1220)  & 0.036 (1270)  \\
\hline
$P_{\rm e}$\tablenotemark{i}                    & 44.44        & 44.45       & 44.85   & 44.57      \\
$P_{\rm B}$\tablenotemark{j}                    & 44.00        & 44.68       & 44.80   & 44.92      \\
$P_{\rm r}$\tablenotemark{k}                    & 44.86        & 45.63       & 46.42   & 45.49      \\
$P_{\rm p}$\tablenotemark{l}                    & 46.48        & 46.48       & 46.86   & 46.97      \\
\tableline
\end{tabular}
\tablenotetext{1}{Slope of particle spectral index before break energy.}
\tablenotetext{2}{Slope of particle spectral index after break energy.}
\tablenotetext{3}{Magnetic field in Gauss.}
\tablenotetext{4}{Particle energy density in erg cm$^{-3}$.}
\tablenotetext{5}{Bulk Lorentz factor.}
\tablenotetext{6}{Break Lorentz factor of electrons.}
\tablenotetext{7}{Maximum Lorentz factor of electrons.}
\tablenotetext{8}{Distance of the emission region from central black hole in parsec (in $R_{\rm Sch}$).}
\tablenotetext{9}{Jet power in electrons in log scale.}
\tablenotetext{10}{Magnetic jet power in log scale.}
\tablenotetext{11}{Radiative jet power in log scale.}
\tablenotetext{12}{Jet power in protons in log scale.}
\end{center}
}
\end{table*}

\begin{table}
\caption{Flare characteristics obtained by fitting the three flares seen during the Fermi ToO monitoring (Figure~\ref{fig:flare_profile}). Errors are estimated at 1$\sigma$ level.}\label{tab:flare_profile}
 \begin{tabular}{ccccccc}
\hline \hline
Name & $F_c$ & $F_p$ & $t_p$ & $T_r$ & $T_f$ & $\chi^2_r$ \\
\hline
F1 & 0.98 $\pm$ 0.13 &  3.73 $\pm$ 1.41 & 56749.55 $\pm$ 0.04 & 0.033 $\pm$ 0.026 & 0.268 $\pm$ 0.153 & 0.47 \\
F2 & 1.61 $\pm$ 0.70 & 18.78 $\pm$ 2.56 & 56750.25 $\pm$ 0.04 & 0.078 $\pm$ 0.026 & 0.295 $\pm$ 0.062 & 1.71 \\
F3 & 3.30 $\pm$ 0.24 & 10.93 $\pm$ 6.17 & 56751.26 $\pm$ 0.09 & 0.085 $\pm$ 0.042 & 0.038 $\pm$ 0.073 & 0.75 \\

\hline
\end{tabular}
\tablecomments{Fluxes $F_c$ and $F_p$ are in 10$^{-6}$ \phflux, $t_p$ has the unit of MJD, and $T_r$ and $T_f$ are in days.}
\end{table}

\begin{table*}
{\small
\begin{center}
\caption{Summary of the parameters used/derived from the modeling of the SEDs in Figure~\ref{fig:sed_all}. Symbols have their usual meanings as in Table~\ref{tab:sed}.}\label{tab:sed_1day}
\begin{tabular}{lcccccc}
\tableline
\tableline
Parameter           &  Day 1       & Day 2        & Day 3         & Day 4        & Day 5        & Day 6       \\
\tableline
$p$                 & 1.5          & 1.7          & 1.7           & 1.7          & 1.7          & 1.7         \\
$q$                 & 4.4          & 4.5          & 4.5           & 4.5          & 4.5          & 4.5         \\
$B$                 & 4.6          & 2.5          & 2.5           & 2.5          & 2.5          & 2.7         \\
$U'_{\rm e}$        & 0.09         & 0.19         & 0.18          & 0.18         & 0.17         & 0.18        \\
$\Gamma$            & 18           & 22           & 20            & 22           & 18           & 20          \\
$\gamma'_{\rm b}$   & 336          & 637          & 634           & 637          & 635          & 526         \\
$\gamma'_{max}$     & 3e4          & 3e4          & 3e4           & 3e4          & 3e4          & 3e4         \\
$R_{\rm diss}$      & 0.034 (1170) & 0.034 (1170) & 0.034 (1170)  & 0.034 (1170) & 0.034 (1180) & 0.034 (1180)\\
\hline
$P_{\rm e}$         & 44.45        & 44.97        & 44.85         & 44.95        & 44.76        & 44.87       \\
$P_{\rm B}$         & 45.44        & 45.08        & 45.00         & 45.09        & 44.92        & 45.08       \\
$P_{\rm r}$         & 45.97        & 46.79        & 46.52         & 46.77        & 46.26        & 46.47       \\
$P_{\rm p}$         & 46.43        & 47.07        & 46.96         & 47.05        & 46.86        & 47.00       \\
\tableline
\end{tabular}
\end{center}
}
\end{table*}

\newpage

\newpage

\newpage

\newpage

\newpage

\newpage

\newpage

\newpage

\end{document}